\documentclass[11pt]{article}

\def\ds{\displaystyle}

\makeatletter

\topmargin\z@

\def\section{\@startsection{section}{1}{\z@}{-\bigskipamount}{\bigskipamount}
{\bf}}

\newcommand{\bfm}[1]{\mathbf{#1}}

\newfont{\tts}{cmr5 scaled 1000}

\font\Bbb=msbm10 scaled\magstep1
\DeclareSymbolFont{AMSa}{U}{msa}{m}{n}
\DeclareSymbolFont{AMSb}{U}{msb}{m}{n}
\DeclareSymbolFontAlphabet{\mathbb}{AMSb}
\DeclareMathSymbol\square           {\mathord}{AMSa}{"03}

\def\Bbb{\mathbb}

\def\scsc{\scriptscriptstyle}
\def\ds{\displaystyle}

\def\C{{\Bbb{C}}}
\def\N{{\Bbb{N}}}
\def\R{{\Bbb{R}}}

\begin{document}

\input prepictex
\input pictex
\input postpictex

\renewcommand\footnoterule{\noindent\rule{1.6cm}{0.025cm}\medskip}

\thispagestyle{empty}

\begin{center}
{\Large\bf  Discrete phase space - II:\\
The second quantization of free\\[0.2cm]
relativistic wave fields} \\[1.2cm]
{\large\bf A. Das\footnote[1]{{\it E-mail address:} das@sfu.ca}}
\end{center}
\vskip0.6cm

\centerline{\it Department of Mathematics and Statistics}
\centerline{\it Simon Fraser University, Burnaby, B.C.,
V5A 1S6, Canada}

\vskip0.6cm

\noindent
\hrulefill
\vskip0.2cm

\renewcommand{\baselinestretch}{0.9}
\small\normalsize

\noindent
{\small\bf Abstract}\bigskip

\noindent
{\small
The Klein-Gordon equation, the Maxwell equation, and the Dirac
equation are presented as partial difference equations in the {\em
eight-dimensional covariant discrete phase space}. These equations
are also furnished as {\em difference-differential} equations in
the arena of {\em discrete phase space and continuous time.} The
scalar field and electromagnetic fields are quantized with commutation
relations. The spin-1/2 field is quantized with anti-commutation
relations. Moreover, the total momentum, energy and charge of these
free relativisitic quantized fields in the discrete phase space and
continuous time are computed {\em exactly}. The results agree
completely with those computed from the relativisitic fields defined
on the space-time continuum.
}
\vskip0.6cm


\noindent
\hrulefill
\vskip1cm

\renewcommand{\baselinestretch}{1}
\small\normalsize

\section{Introduction}

In the preceding paper [1], we have presented the
Lagrangian formalism for the relativistic wave fields in the covariant
discrete phase space, as well as in the discrete phase space and
continuous time. In this paper, we shall choose {\em three special\/}
examples of relativistic fields [2]. These are the free scalar field,
electromagnetic fields, and the Dirac spin-1/2 field. Moreover, we
shall quantize these fields with the usual commutation and
anti-commutation rules. We generalize variational techniques for the
operator-valued second-quantized wave fields. The {\em non-singular\/}
Green's functions for various {\em difference\/} equations are provided
{[2,3]}. Finally, we compute the totally conserved four-momentum and
charge for three different fields exactly. These computations are
performed for the wave-equations satisfying {\em
difference-differential equations in the discrete phase space and
continuous time only.} (We do {\em not\/} consider wave fields in the
covariant discrete phase space to calculate the totally conserved
quantities for some physical reasons.) The results of these
computations are {\em identical\/} to those calculated from the usual
relativistic quantum theory of the free fields in the Minkowski
space-time. One may wonder about the utility of such a complicated,
alternate formulation of the quantum theory of free fields! Actually,
the present treatment of the free quantized fields in the discrete
phase space and continuous time is just a {\em prelude\/} to the more
exciting investigations of the interacting fields and the $S$-matrix
in the following paper III.
\vskip1.2cm

\section{Definitions and notations}

We use physical units such that $h=c=\ell=1.$ (Here, $\ell$
is a characteristic length.) All physical quantities are expressed
as dimensionless numbers. Greek indices take from $\{1,2,3,4\}$ and roman
indices take from $\{1,2,3\}$. The summation convention is followed. The
Minkowski metric is denoted by $\eta_{\mu\nu}$ and the signature of
the metric is $+2.$ We denote the set of all real numbers by $\R$ and
all non-negative integers by $\N.$ A bold roman letter indicates a
three-dimensional vector. The equations in the covariant discrete
phase space are denoted by $(..A),$ whereas the equations in the
discrete phase space and continuous time are labelled by $(..B).$

An integer $n^\mu$ is associated with a phase space circle of radius
$\sqrt{2n^\mu+1}$ for each $\mu \in \{1,2,3,4\}.$ Let a function be
defined by $f: \,\N^4 \rightarrow \R$ (or $\N^4 \rightarrow \C$).
The right-difference, the left-difference, and the weighted mean
difference are defined respectively \vspace*{0.2cm}by:
$$
\Delta_\mu f(n) := f(..,n^\mu+1,..) - f(..,n^\mu,..)\vspace*{0.1cm}\,,
\eqno{\rm (1i)}
$$
$$
\Delta_\mu^{\prime} f(n) := f(..,n^\mu,..) -
f(..,n^\mu-1,..)\vspace*{0.25cm}\,,
\eqno{\rm (1ii)}
$$
$$
\Delta_\mu^{\scsc\#} f(n) := (1 / \sqrt{2}\,) \Big[ \sqrt{n^\mu+1}
\,{f}(..,n^\mu+1,..) - \sqrt{n^\mu}
\,f(..,n^\mu-1,..)\Big]\vspace*{0.4cm}\,.
\eqno{\rm (1iii)}
$$
Let us work out an example to elaborate the definition (1iii).
Suppose that $k_1$ is a real number and $H_{n^1}(k_1)$ is a Hermite
polynomial. (See Appendix~I.)  In that \vspace*{0.2cm}case
$$
\begin{array}{rrl}
\Delta _1^{\scsc\#} \xi_{n^1} (k_1) &:=& \ds \Delta_1^{\scsc\#}
\Biggr[ \frac{(i)^{n^1} e^{-(k_1)^2/2} H_{n^1}(k_1)}{\pi^{1/4}
2^{n^1/2} \sqrt{(n^1)}!}\Biggr] \\[0.8cm]
&=& \ds \frac{i^{(n^1+1)} e^{-(k_1)^2/2}}{2\pi^{1/4} 2^{n^1/2}
\sqrt{(n^1)}!} \,\Big[H_{n^1+1} (k_1) + 2n_1 H_{n^1-1} (k_1)\Big]
\\[0.8cm]
&=& \ds (ik_1) \,\Biggr[\frac{(i)^{n^1} e^{-(k_1)^2/2} H_{n^1}
(k_1)}{\pi^{1/4} 2^{n^1/2} \sqrt{(n^1)}!}\Biggr] = (ik_1) \xi_{n^1}
(k_1)\vspace*{0.5cm}\,.
\end{array}
\eqno(2)
$$

In the case of $\phi (n) \equiv \phi (n^1,n^2,n^3,n^4)$ is an
operator-valued function over a domain of $\N^4,$ we adopt the {\em
same\/} definitions as in (1i,ii,iii).
\vskip1.2cm

\section{The second quantization of a free non-hermitian scalar field}

Let $\rho, \rho_\mu$ be five linear operators over a
non-separable Hilbert space [4]. Let $\rho^{\dagger},
\rho_\mu^{\dagger}$ denote the corresponding adjoint or
hermitian-conjugate operators. The linear operator-valued Lagrangian
function $L$ of ten operators is defined to be [5]:
$$
\begin{array}{l}
L(\rho,\rho^{\dagger}; \rho_\mu , \rho_\mu^{\dagger}) :=
-(\eta^{\mu\nu} \rho_\mu^{\dagger} \rho_\nu + \mu^{2} \rho^{\dagger}
\rho)\,, \\[0.5cm]
{\ds \frac{\partial L(..)}{\partial \rho} = -\mu^2 \rho^{\dagger}\,,
\; \frac{\partial L(..)}{\partial \rho^{\dagger}} = -\mu^2 \rho\,,}
\\[0.7cm]
{\ds \frac{\partial L(..)}{\partial \rho_\mu} = -\eta^{\mu\nu}
\rho_\nu^{\dagger}\,, \;
\frac{\partial L(..)}{\partial \rho_\mu^{\dagger}} = -\eta^{\mu\nu}
\rho_{\nu}\,,} \\[0.7cm]
{\ds \frac{\partial^2 L(..)}{\partial \rho \partial \rho^{\dagger}} =
\frac{\partial^2 L(..)}{\partial \rho^{\dagger} \partial \rho} =
-\mu^2\,{\rm I}\,, \,\; \frac{\partial^2 L(..)}{(\partial \rho)^2} =
\frac{\partial^2 L(..)}{(\partial \rho^{\dagger})^2} \equiv 0\,,}
\\[0.7cm]
{\ds \frac{\partial^2 L(..)}{\partial \rho_\mu^{\dagger} \partial
\rho_{\nu}} = \frac{\partial^2 L(..)}{\partial \rho_{\nu}
\partial \rho_\mu^{\dagger}} = -\eta^{\mu\nu}\,{\rm I}\,, \,\;
\frac{\partial^2 L(..)}{\partial \rho_\mu \partial \rho_\nu} =
\frac{\partial^2 L(..)}{\partial \rho_\mu^{\dagger} \partial
\rho_\nu^{\dagger}} \equiv 0\,. }
\end{array}
\eqno(3)
$$
\vskip0.4cm

\noindent
Here, $\mu > 0$ is the mass parameter, ``I'' is the identity operator,
and ``O'' is the zero operator. The linear operators do {\em not\/}
commute necessarily. Therefore, the {\em order\/} in which they appear
must be {\em preserved}. The $j$-th partial derivatives of $L$ in
(3) with respect to $\rho, \rho ^{\dagger}, \rho_\mu,
\rho_\mu^{\dagger}$ are all zero operators for $j\geq 3.$ Assuming
that the Euler-Lagrange equations $(38\overline{\rm A}, \overline{\rm
B})$ of paper~I  are valid for the operator-valued functions $\rho
= \phi (n)$ and $\rho = \phi ({\bfm{n}},t),$ we obtain [2] from
\vspace*{0.2cm}(3)
$$
\eta^{\mu\nu} \Delta_\mu^{\scsc\#} \Delta_\nu^{\scsc\#} \phi (n) -
\mu^2 \phi (n) = 0\vspace*{0.2cm}\,,
\eqno{\rm (4A)}
$$
$$
\delta^{ab} \Delta_a^{\scsc\#} \Delta_b^{\scsc\#} \phi ({\bfm{n}},t)
- (\partial_t)^2 \phi ({\bfm{n}},t) - \mu^2 \phi ({\bfm{n}},t) =
0\vspace*{0.3cm}\,.
\eqno{\rm (4B)}
$$
(Strictly speaking, $\phi$ is a section of the fibre bundle [6] of
linear operators over the base ``manifold'' $\N^4$ or $\N^3\times
\R.$)

Now we shall derive the operator difference conservation equations.
Adapting the equations (A.II.2), (A.II.3), and (A.II.4A) of paper~I
to the\linebreak Lagrangian in (3) and noting that operator
$\frac{\partial^2 L(..)}{\partial \rho^{\dagger} \partial \rho}$
etc. are proportional to the identity, we obtain relativistic
conservation equations\vspace*{0.2cm}:
$$
\begin{array}{l}
{\ds \left\{ \Delta_\nu^{\scsc\#} \left[\frac{\partial L(..)}{\partial
\rho_\nu}\right]_{|..} \cdot \Delta_\mu^{\scsc\#} \phi +
\frac{\partial L(..)}{\partial \rho_{\nu |..}} \cdot
\Delta_\mu^{\scsc\#} \Delta_\nu^{\scsc\#} \phi + {\rm (h.c.)}
- \Delta_\mu^{\scsc\#} L(..)_{|..} \right\} } \\[0.7cm]
{\ds + \frac{1}{2} \left[\frac{\partial^2 L(..)}{\partial\rho^{\dagger}
\partial \rho} + \frac{\partial^2 L(..)}{\partial \rho \partial
\rho^{\dagger}}\right]_{|..} \cdot \Big\{\Delta_\mu^{\scsc\#}
\Big[\phi^{\dagger}(n) \cdot \phi (n)\Big] } \\[0.6cm]
{\ds - \phi^{\dagger} (n) \cdot
\Delta_\mu^{\scsc\#} \phi - \big[\Delta_\mu^{\scsc\#}
\phi^{\dagger}\big] \cdot \phi (n)\Big\}}
\end{array}
$$
$$
\begin{array}{l}
{\ds + \frac{1}{2} \left[\frac{\partial^2 L(..)}{\partial
\rho_\sigma^{\dagger} \partial \rho_\nu} + \frac{\partial^2
L(..)}{\partial \rho_\sigma \partial \rho_\nu^{\dagger}}\right]_{|..}
\cdot \Big\{\Delta_\mu^{\scsc\#} \Big[\Delta_\nu^{\scsc\#} \phi^{\dagger}
\cdot \Delta_\sigma^{\scsc\#} \phi\Big] } \\[0.7cm]
{\ds - \Delta_\nu^{\scsc\#}\phi^{\dagger} \cdot
\Delta_\mu^{\scsc\#} \Delta_\sigma^{\scsc\#} \phi
- \Big[\Delta_\mu^{\scsc\#} \Delta_\nu^{\scsc\#} \phi^{\dagger}\Big]
\cdot \Delta_\sigma^{\scsc\#} \phi\Big\}} \\[0.5cm]
{\ds + \Big[\Delta_\mu^{\scsc\#}(1)\Big] \cdot \Biggr\{ L(..)_{|..}
- \frac{\partial L(..)}{\partial \rho_{|..}} \cdot \phi (n)} \\[0.7cm]
{\ds - \phi^{\dagger} (n) \cdot \frac{\partial L(..)}{\partial
\rho_{|..}^{\dagger}} - \frac{\partial L(..)}{\partial \rho_{\nu |..}}
\cdot \Delta_\nu^{\scsc\#} \phi - \Delta_\nu^{\scsc\#} \phi^{\dagger}
\cdot \frac{\partial L(..)}{\partial \rho_\nu^{\dagger}} } \\[0.8cm]
{\ds + \frac{1}{2} \left[\frac{\partial^2 L(..)}{\partial
\rho^{\dagger} \partial \rho} + \frac{\partial^2 L(..)}{\partial \rho
\partial \rho^{\dagger}}\right]_{|..} \cdot \phi^{\dagger}(n) \phi(n)
} \\[0.8cm]
{\ds + \frac{1}{2} \left[\frac{\partial^2 L(..)}{\partial
\rho_\sigma^{\dagger}
\partial \rho_\nu} + \frac{\partial^2 L(..)}{\partial \rho_\sigma
\partial \rho_\nu^{\dagger}}\right]_{|..} \cdot \Delta_\nu^{\scsc\#}
\phi^{\dagger} \cdot \Delta_\sigma^{\scsc\#} \phi \Biggr\}
= 0\,. }
\end{array}
\eqno{\rm (5A)}
$$
\vskip0.4cm

\noindent
Here, (h.c.) stands for the hermitian-conjugation of the preceding
terms.

Since the Lagrangian function $L$ is a second degree polynomial, the
relativistic conservation equation (5A) is {\em exact\/} and no
additional terms indicated by $\ldots$ are necessary.

Moreover, the last curly bracket in (5A) is exactly the {\em zero\/}
operator. Using (5v) of paper~I  and (6i), after a long
calculation, the relativisitic equation (5A) yields the {\em exact\/}
conservation equation\vspace*{0.4cm}:
$$
\hspace*{-0.35cm}\begin{array}{l}
{\ds \Delta_\nu \Biggr\{\sqrt{\frac{n^\nu}{2}} \Biggr[\frac{\partial
L(..)}{\partial \rho_{\nu |(..,n^\nu -1,..)}}\Biggr] \cdot
\Delta_\mu^{\scsc\#}
\phi + \frac{\partial L(..)}{\partial_{\nu |..}} \cdot
\big[\Delta_\mu^{\scsc\#} \phi \big]_{|(..,n^\nu-1,..)}
\!+\!{\mathrm{(h.c.)}} } \\[0.8cm]
{\ds -\delta_\mu^\nu \biggr[-\eta^{\rho\sigma} \Big(\Delta_\rho^{\scsc\#}
\phi^{\dagger}_{|(..,n^\nu-1,..)} \cdot \Delta_\sigma^{\scsc\#} \phi
+ \Delta_\rho^{\scsc\#} \phi^{\dagger} \cdot \big[\Delta_\sigma^{\scsc\#}
\phi \big]_{|(..,n^\nu-1,..)}\Big) } \\[0.6cm]
{\ds -\mu^2 \Big(\phi^{\dagger}(..,n^\nu-1,..) \cdot \phi (n)
+ \phi^{\dagger} (n) \cdot \phi (..,n^\nu-1,..)\Big)\biggr]\Biggr\}
= 0\,. }
\end{array}
\eqno{\rm (6A)}
$$
\vskip0.3cm

\noindent
In a similar fashion, we can derive {\em exact\/}
difference-differential equations (which are equivalent to the
relativistic equations)\vspace*{0.4cm}:
$$
\Delta_b T_a^b + \partial_t T_a^4 = 0\vspace*{0.3cm}\,,
\eqno{\rm (6Bi)}
$$
$$
\Delta_b T_4^b + \partial_t T_4^4 = 0\,,
\eqno{\rm (6Bii)}
$$

$$
\begin{array}{l}
{\ds T_a^b ({\bfm{n}},t) := \sqrt{\frac{n^a}{2}} \left\{\frac{\partial
L(..)}{\partial \rho_{b|(..,n^b-1,..)}}\right\} \cdot \Delta_a^{\scsc\#}
\phi } \\[0.4cm]
{\ds + \frac{\partial L(..)}{\partial \rho_{b|..}} \cdot
\big[\Delta_a^{\scsc\#} \phi \big]_{|(..,n^b-1,..)} + {\mathrm{(h.c.)}}
} \\[0.6cm]
{\ds -\delta_a^b \Big[\big(\Delta_c^{\scsc\#}
\phi^{\dagger}\big)_{|(..,n^b-1,..)} \cdot \Delta_d^{\scsc\#} \phi
+ \Delta_c^{\scsc\#} \phi^{\dagger} \cdot \big(\Delta_d^{\scsc\#}
\phi\big)_{|(..,n^b-1,..)}\Big] } \\[0.4cm]
{\ds + \Big[\big(\partial_t \phi^{\dagger}\big)_{|(..,n^b-1,..)} \cdot
\partial_t \phi + \partial_t \phi^{\dagger} \cdot \big(\partial_t
\phi \big)_{|(..,n^b-1,..)}\Big] } \\[0.4cm]
{\ds -\mu^2 \Big[\phi^{\dagger} (..,n^b-1,..) \cdot \phi (n) +
\phi^{\dagger}(n) \cdot \phi (..,n^b-1,..)\Big]\,,}
\end{array}
\eqno{\rm (6Biii)}
$$
\vskip0.3cm

$$
T_a^4 ({\bfm{n}},t) := \frac{\partial L(..)}{\partial
\rho_{4|..}} \Delta_a^{\scsc\#} \phi + \Delta_a^{\scsc\#}
\phi^{\dagger} \cdot \frac{\partial L(..)}{\partial
\rho_{4|..}^{\dagger}}\vspace*{0.2cm}\,,
\eqno{\rm (6Biv)}
$$
$$
T_4^a ({\bfm{n}},t) := \sqrt{\frac{n^a}{2}} \left\{ \frac{\partial
L(..)}{\partial \rho_{a|..}} \cdot \partial_t \phi + {\mathrm{(h.c.)}}
\right\}\vspace*{0.2cm}\,,
\eqno{\rm (6Bv)}
$$
$$
T_4^4 ({\bfm{n}},t) := \left\{ \frac{\partial L(..)}{\partial
\rho_{4|..}} \cdot \partial_t \phi + {\mathrm{(h.c.)}} - L(..)_{|..}
\right\}\,.
\eqno{\rm (6Bvi)}
$$
\vskip0.5cm

It is instructive to compare and contrast the equations
(39Bi,ii,iii,iv,v,vi) of paper~I with the equations
(6Bi,ii,iii,iv,v,vi). The first set of $T_\nu^\mu ({\bfm{n}},t)$
satisfy possibly approximate conservation equations whereas the second
set satisfy {\em exact\/} equations. Neither of these $T_\nu^\mu
({\bfm{n}},t)$ obey exactly the tensorial transformation rules
(34B). However, the {\em relativistic\/} total four-momentum
components $P_\mu\!$'s and the invariant total charge $Q$ can be
elicited from
$$
P_b = -\sum\limits_{{\bfm{n}}=0}^{\infty (3)} \left\{ \frac{\partial
L(..)}{\partial \rho_{4|..}} \cdot \Delta_b^{\scsc\#} \phi +
\Delta_b^{\scsc\#} \phi^{\dagger} \cdot \frac{\partial L(..)}{\partial
\rho_{4|..}^{\dagger}} \right\}_{\!\!|t=0} \vspace*{0.3cm}\,,
\eqno{\rm (7Bi)}
$$
$$
H = -P_4 = \sum\limits_{{\bfm{n}}=0}^{\infty (3)} \left\{ \frac{\partial
L(..)}{\partial \rho_{4|..}} \cdot \partial_t \phi + \partial_t
\phi^{\dagger} \cdot \frac{\partial L(..)}{\partial
\rho_{4|..}^{\dagger}}-L(..)_{|..}\right\}_{\!\!|t=0}\vspace*{0.5cm}\,,
\eqno{\rm (7Bii)}
$$
$$
Q = -ie \sum\limits_{{\bfm{n}}=0}^{\infty (3)} \left\{ \frac{\partial
L(..)}{\partial \rho_{4|..}} \cdot \phi ({\bfm{n}},t) - \phi^{\dagger}
({\bfm{n}},t) \cdot \frac{\partial L(..)}{\partial
\rho_{4|..}^{\dagger}} \right\}_{\!\!|t=0} \,.
\eqno{\rm (7Biii)}
$$
\vskip0.3cm

\noindent
Note that above {\em relativistic\/} equations are {\em exact\/} and
no additional terms denoted by $\ldots$ are necessary. We furnish
a general class of exact solutions [2] (``the plane wave
superposition'') of the (generalized) Klein-Gordon equations (4A,B)
in the following:
$$
\begin{array}{rcl}
\phi (n) &=& {\ds \int\limits_{\R^3} [2\omega ({\bfm{k}})]^{-1/2} \left\{
a({\bfm{k}}) \left[\,\prod\limits_{\mu=1}^4 \xi_{n^\mu} (k_\mu)\right]
\right. } \\[0.6cm]
&+& {\ds \left. b^{\dagger} ({\bfm{k}}) \left[\,\prod\limits_{\mu=1}^4
\,\overline{\xi_{n^\mu}(k_\mu)}\right]\right\} d^3 {\bfm{k}}
} \\[0.6cm]
&=&: \phi^{-}(n) + \phi^{+}(n)\,,
\end{array}
\eqno{\rm (8A)}
$$
\vskip0.5cm

$$
\begin{array}{rcl}
\phi ({\bfm{n}},t) &=& {\ds \int\limits_{\R^3} [2\omega
({\bfm{k}})]^{-1/2} \left\{ a({\bfm{k}}) \left[\,\prod\limits_{j=1}^3
\xi_{n^j} (k_j)\right] e^{-i\omega t} \right. } \\[0.6cm]
&+& {\ds \left. b^{\dagger} ({\bfm{k}}) \left[\,\prod\limits_{j=1}^3
\,\overline{\xi_{n^j}(k_j)}\right] e^{i\omega t}\right\} d^3 {\bfm{k}}
} \\[0.6cm]
&=:& \phi^{-}({\bfm{n}},t) + \phi^{+}({\bfm{n}},t)\,,
\end{array}
\eqno{\rm (8B)}
$$
\vskip0.6cm

$$
({\bfm{k}}) := (k_1,k_2,k_3), \;d^3 {\bfm{k}} := dk_1, dk_2,
dk_3\vspace*{0.4cm}\,,
\eqno{\rm (9i)}
$$
$$
-k_4 \equiv \omega ({\bfm{k}}) := + \sqrt{(k_1)^2 + (k_2)^2 + (k_3)^2
+ \mu^2} > 0\vspace*{0.3cm}\,,
\eqno{\rm (9ii)}
$$
$$
\xi_{n^\mu} (k_\mu) := (i)^{n^\mu} \:\frac{\exp [-(k_\mu)^2/2]\cdot
H_{n^\mu}(k_\mu)}{(\pi)^{1/4}\, 2^{n^\mu /2} \sqrt{(n^\mu)!}}\,.
\eqno{\rm (9iii)}
$$
\vskip0.5cm

\noindent
Here, the indices $\mu$ and $j$ are {\em not\/} summed. Moreover,
$H_{n^\mu} (k_\mu)$ stands for a Hermite polynomial. (For the
properties of orthonormal complex-polynomials $\xi_{n^\mu} (k_\mu),$
see equation (2) and Appendix~I.) The functions ``$a$'' and ``$b$''
are some sections of the fibre-bundle [6] of the linear operators
over the base manifold $\R^3$ (the momentum-space). The
operator-valued improper \vspace*{0.075cm}integrals
$$
\int\limits_{\R^3}\!a^{\dagger} ({\bfm{k}})a ({\bfm{k}}) d^3 {\bfm{k}},
\; \int\limits_{\R^3}\!a ({\bfm{k}}) a^{\dagger} ({\bfm{k}}) d^3
{\bfm{k}}, \; \int\limits_{\R^3}\!b^{\dagger} ({\bfm{k}})
b ({\bfm{k}}) d^3 {\bfm{k}}, \; \int\limits_{\R^3}\!b ({\bfm{k}})
b^{\dagger} ({\bfm{k}}) d^3 {\bfm{k}}, \;\; {\rm etc.}
\vspace*{0.075cm}\,
$$
should converge in certain sense for the existence of (8A,B). There
are more restrictions on these operators which follow from the quantum
theory. These are the following canonical quantum rules to be imposed
on the operators $a, a^{\dagger}, b, b^{\dagger}$:

$$
\begin{array}{l}
[A,B] := AB - BA\,, \\
\delta^3 ({\bfm{k}}-\widehat{\bfm{k}}) := {\ds \delta
(k_1-\widehat{k}_1) \,\delta (k_2-\widehat{k}_2) \,\delta
(k_3-\widehat{k}_3) = \prod\limits_{j=1}^{3} \delta (k_j-\widehat{k}_j)
\,, } \\
{[a({\bfm{k}}), a^{\dagger}(\widehat{\bfm{k}})]} = [b(\widehat{\bfm{k}}),
b^{\dagger}(\widehat{\bfm{k}})] = \delta^3 ({\bfm{k}}-\widehat{\bfm{k}})
\,{\mathrm I}\,({\bfm{k}})\,, \\[0.3cm]
{[a({\bfm{k}}), a(\widehat{\bfm{k}})]} = [a^{\dagger}({\bfm{k}}),
a^{\dagger}(\widehat{\bfm{k}})] = [b({\bfm{k}}), b(\widehat{\bfm{k}})]
= [b^{\dagger} ({\bfm{k}}), b^{\dagger}(\widehat{\bfm{k}})] = 0\,.
\end{array}
\eqno(10)
$$

\noindent
Here, $\delta (k_j-\widehat{k}_j)$ denotes a Dirac-delta distribution
function, ${\rm I}\,({\bfm{k}})$ stands for the identity operator, and
``0'' denotes the zero operator. The linear operators $a({\bfm{k}}),
a^{\dagger} ({\bfm{k}})$ are called the destruction and creation
operators for particles (or field quantas). The particle and
anti-particle vacuum is denoted by the Hilbert vector $|\psi_0\rangle
.$ The particle and anti-particle (occupation) number operators
$N^+({\bfm{k}}), N^{-}({\bfm{k}})$ are defined by and satisfy the
following equations:
$$
\begin{array}{l}
N^+({\bfm{k}}) := a^{\dagger}({\bfm{k}}) a({\bfm{k}}), \; N^{-}
({\bfm{k}}) := b^{\dagger} ({\bfm{k}}) b ({\bfm{k}})\,, \\[0.25cm]
a ({\bfm{k}}) |\psi_0\rangle = b({\bfm{k}}) |\psi_0\rangle = N^+
({\bfm{k}}) |\psi_0\rangle = N^{-} ({\bfm{k}}) |\psi_0\rangle =
|{\bfm{0}}\rangle\,, \\[0.25cm]
\langle \psi_0|\psi_0\rangle = 1, \;\,\langle {\bfm{0}}|{\bfm{0}}\rangle
= 0\,, \\[0.25cm]
{[N^+ ({\bfm{k}}), a(\widehat{\bfm{k}})]} = -\delta^3
({\bfm{k}}-\widehat{\bfm{k}}) a({\bfm{k}})\,, \\[0.25cm]
{[N^+ ({\bfm{k}}), a^{\dagger}(\widehat{\bfm{k}})]} = \delta^3
({\bfm{k}}-\widehat{\bfm{k}}) a^{\dagger}({\bfm{k}})\,, \\[0.25cm]
{[N^- ({\bfm{k}}), b(\widehat{\bfm{k}})]} = -\delta^3
({\bfm{k}}-\widehat{\bfm{k}}) b({\bfm{k}})\,, \\[0.25cm]
{[N^- ({\bfm{k}}), b^{\dagger}(\widehat{\bfm{k}})]} = \delta^3
({\bfm{k}}-\widehat{\bfm{k}}) b^{\dagger}({\bfm{k}})\,, \\[0.25cm]
{[N^+ ({\bfm{k}}), b(\widehat{\bfm{k}})]} = [N^+({\bfm{k}}), b^{\dagger}
(\widehat{\bfm{k}})] \equiv 0\,, \\[0.25cm]
{[N^- ({\bfm{k}}), a(\widehat{\bfm{k}})]} = [N^-({\bfm{k}}), a^{\dagger}
(\widehat{\bfm{k}})] \equiv 0\,, \\[0.25cm]
{[N^+ ({\bfm{k}}), N^-(\widehat{\bfm{k}})]} \equiv 0\,.
\end{array}
\eqno(11)
$$
\vskip0.1cm

\noindent
The eigenvalues of $N^+({\bfm{k}})$ and $N^-({\bfm{k}}),$ the so
called occupation numbers, take values from $\N := \{0,1,2,3,\ldots
\}.$ Therefore, the particles and anti-particles of the quantized
scalar field obey the Bose-Einstein statistics.

The covariant commutation relations which follow from (8A) and (10)
are the following (see Appendix II):
$$
\begin{array}{l}
{[\phi^-(n), \phi^-(\widehat{n})]} = [\phi^+(n), \phi^+(\widehat{n})]
= \big[ (\phi^-(n))^{\dagger}, (\phi^-(\widehat{n}))^{\dagger}\big]
\\[0.25cm]
\phantom{[\phi^-(n), \phi^-(\widehat{n})]}
= \big[ (\phi^+(n))^{\dagger}, (\phi^+(\widehat{n}))^{\dagger}\big]
\equiv 0\,, \\[0.25cm]
{[\phi^-(n), (\phi^-(\widehat{n}))^{\dagger}]} = (i/2\pi )\Delta_+
(n,\widehat{n}; \mu)\,{\mathrm{I}}\,, \\[0.25cm]
{[\phi^+(n), (\phi^+(\widehat{n}))^{\dagger}]} = (i/2\pi )\Delta_{-}
(n,\widehat{n}; \mu)\,{\mathrm{I}}\,, \\[0.25cm]
{[\phi (n), \phi (\widehat{n})]} = \big[(\phi (n))^{\dagger}, (\phi
(\widehat{n}))^{\dagger}\big] \equiv 0\,, \\[0.25cm]
{[\phi(n), (\phi (\widehat{n}))^{\dagger}]} = (i/2\pi )\Delta
(n,\widehat{n}; \mu)\,{\mathrm{I}}\,.
\end{array}
\eqno{\rm (12A)}
$$

Moreover, the covariant commutation relations in the
difference-differential representation are:
$$
\hspace*{-0.325cm}\begin{array}{l}
[\phi^{-}({\bfm{n}},t), \phi^{-}(\widehat{\bfm{n}},\widehat{t}\,)]
= [\phi^+({\bfm{n}},t), \phi^+(\widehat{\bfm{n}},\widehat{t}\,)]
= \big[(\phi^{-}({\bfm{n}},t))^{\dagger}, (\phi^{-}(\widehat{\bfm{n}},
\widehat{t}\,))^{\dagger}\big] \\[0.25cm]
\phantom{[\phi^{-}({\bfm{n}},t),
\phi^{-}(\widehat{\bfm{n}},\widehat{t}\,)]}
= \big[(\phi^+ ({\bfm{n}},t))^{\dagger}, (\phi^+
(\widehat{\bfm{n}}, \widehat{t}\,))^{\dagger}\big] \equiv 0\,,
\\[0.25cm]
{[\phi^{-}({\bfm{n}},t), (\phi^{-}(\widehat{\bfm{n}},
\widehat{t}\,))^{\dagger}]} = i\Delta_+ ({\bfm{n}},t;\,\widehat{\bfm{n}},
\widehat{t}; \,\mu )\,{\mathrm{I}}\,, \\[0.25cm]
{[\phi^{+}({\bfm{n}},t), (\phi^{+}(\widehat{\bfm{n}},
\widehat{t}\,))^{\dagger}]} = i\Delta_{-} ({\bfm{n}},t;\,\widehat{\bfm{n}},
\widehat{t};\,\mu )\,{\mathrm{I}}\,, \\[0.25cm]
{[\phi ({\bfm{n}},t), \phi (\widehat{\bfm{n}},\widehat{t}\,)]}
= \big[\phi ({\bfm{n}},t)^{\dagger}, (\phi
(\widehat{\bfm{n}},\widehat{t}\,))^{\dagger}\big] \equiv 0\,,
\\[0.25cm]
\big[\phi ({\bfm{n}},t), (\phi (\widehat{\bfm{n}},\widehat{t}\,))^{\dagger}
\big] = i\Delta ({\bfm{n}},t;\,\widehat{\bfm{n}},\widehat{t};\,\mu )
\,{\mathrm{I}}\,, \\[0.25cm]
\big[\phi ({\bfm{n}},t), (\phi (\widehat{\bfm{n}},\widehat{t}\,))^{\dagger}
\big]_{|{\hat{t}}=t} \equiv 0 \quad\mbox{for}\quad {\bfm{n}}\neq
\widehat{\bfm{n}}\,, \\[0.15cm]
{\ds [\partial_t \phi ({\bfm{n}},t), (\phi (\widehat{\bfm{n}},
\widehat{t}\,))^{\dagger}]_{|{\hat{t}}=t} = -i
\left(\prod\limits_{j=1}^{3}
\delta_{n^j {\hat{n}}^j}\right) \!{\mathrm{I}} =:
-\delta_{{\bfm{n}}{\hat{\bfm{n}}}}\,{\mathrm{I}}\,,} \\[0.6cm]
\big[\partial_t \phi ({\bfm{n}},t), (\partial_{{\hat{t}}}\,\phi
(\widehat{\bfm{n}}, \widehat{t}\,))^{\dagger}\big]_{|{\hat{t}}=t}
\equiv 0 \quad\mbox{for}\quad
{\bfm{n}}\neq \widehat{\bfm{n}}\,.
\end{array}
\eqno{\rm (12B)}
$$
\vskip0.3cm

\noindent
The last three commutators in (12B) resemble the three fundamental
postulates of quantum mechanics, namely
$$
[Q_a,Q_b] \equiv 0, \quad [P_a,Q_b] = -i\delta_{ab}\,{\mathrm{I}},
\quad [P_a,P_b] \equiv 0\,.
$$
\vskip0.2cm

Now we shall compute the total three-momentum components $P_j$, the
total energy $H,$ and the total charge $Q$ from the equations
(7Bi,ii,iii). (See the comments at the end of Section~5 of paper~I
for not considering equations labelled A.) Elaborate computations
are explicitly performed in Appendix~III. We summarize the results
in the following equations:
$$
P_j = \int\limits_{\R^3} \,[N^+({\bfm{k}}) + N^{-}({\bfm{k}})] \,k_j
d^3 {\bfm{k}}\,,
\eqno{\rm (13i)}
$$
$$
H = -P_4 = \int\limits_{\R^3} \,[N^+ ({\bfm{k}}) + N^{-}({\bfm{k}})
+ \delta^3 (0)\,{\mathrm{I}}\,({\bfm{k}})]\,\omega ({\bfm{k}})
\,d^3 {\bfm{k}}\,,
\eqno({\rm (13ii)}
$$
$$
Q = e \int\limits_{\R^3} \,[N^+ ({\bfm{k}}) - N^{-} ({\bfm{k}})]
\,d^3 {\bfm{k}}\,.
\eqno{\rm (13iii)}
$$
\vskip0.1cm

\noindent
These results are {\em identical\/} to those derived from the usual
relativistic quantum theory of a free non-hermitian scalar field {[7]}
in the (flat) space-time continuum. The equation (13ii) shows that
the divergence of the null-point energy {\em cannot\/} be remedied
by the discrete phase space approach.

We now use the commutators in the equation (10), (11), the field
$\phi$ in (8B), and the conserved operators in (13i,ii,iii) to
derive\vspace*{0.1cm}:
$$
\begin{array}{rcl}
{[P_j, \phi ({\bfm{n}},t)]} &=& -i\Delta_j^{\scsc\#} \phi
({\bfm{n}},t)\,, \\[0.2cm]
{[H, \phi ({\bfm{n}},t)]} &=& -i\partial_t \phi({\bfm{n}},t)\,,
\\[0.2cm]
{[Q, \phi ({\bfm{n}},t)]} &=& -e \phi ({\bfm{n}},t)\,.
\end{array}
\eqno(14)
$$

\noindent
The above equations prove that the operators $P_j, H, Q$ are the {\em
generators\/} for the space translation, the time translation, and
the gauge transformation. (We have put here $\delta^3({\bfm{0}})
0=0.$)

Conservations of the total momentum components $P_j$ and the total
charge $Q$ can be proved alternatively by the commutation relations:
$$
\begin{array}{rcl}
{[H, P_j]} &\equiv & 0\,, \\[0.2cm]
{[H,Q]} &\equiv & 0\,.
\end{array}
\eqno(15)
$$
\vskip1cm

\section{Quantization of free electro-magnetic field}

The Lagrangian function $L$ for the electro-magnetic field is chosen
to be the following second degree polynomial function of the twenty
{[5]} linear, self-adjoint operators $y_\mu$ and
$y_{\mu\nu}$\vspace*{0.1cm}:
$$
\begin{array}{l}
L(y_\mu ; y_{\mu\nu}) := -(1/2) \eta^{\mu\nu} \eta^{\alpha\beta}
y_{\mu\alpha} y_{\nu\beta} \\[0.3cm]
\phantom{L(y_\mu ; y_{\mu\nu})\;}
= -(1/2) [\delta^{ab} \delta^{cd} y_{ac} y_{bd} - \delta^{ab}
(y_{a4} y_{b4} + y_{4a} y_{4b}) + (y_{44})^2]\,, \\[0.4cm]
{\ds \frac{\partial L(..)}{\partial y_\mu} \equiv 0\,, \;\; \frac{\partial
L(..)}{\partial y_{\rho\tau}} = -\eta^{\rho\mu} \eta^{\tau\nu}
y_{\mu\nu}\,, \;\;
\frac{\partial L(..)}{\partial y_{ab}} = -y_{ab}\,, }  \\[0.6cm]
{\ds \frac{\partial L(..)}{\partial y_{4a}} = y_{4a}\,,
\;\; \frac{\partial L(..)}{\partial y_{a4}} = y_{a4}\,, \;\; \frac{\partial
L(..)}{\partial y_{44}} = -y_{44}\,, } \\[0.6cm]
{\ds \frac{\partial^2 L(..)}{(\partial y_{\mu\nu}) (\partial y_{\rho\tau})}
= -\eta^{\rho\mu} \eta^{\tau\nu}\,{\mathrm{I}}\,, } \\[0.6cm]
{\ds \frac{\partial^2 L(..)}{(\partial y_{ab})^2}
= \frac{\partial^2 L(..)}{(\partial
y_{a4})^2} = - \frac{\partial^2 L(..)}{(\partial y_{4a})^2} =
\frac{\partial^2 L(..)}{(\partial y_{44})^2} = -{\mathrm{I}}\,, }
\\[0.6cm]
{\ds \frac{\partial^2 L(..)}{(\partial y_{a4}) (\partial y_{cd})} =
\frac{\partial^2 L(..)}{(\partial y_{4a}) (\partial
y_{cd})} \equiv 0\,. }
\end{array}
\eqno(16)
$$
\vskip0.3cm

\noindent
The third and higher order partial derivatives of $L$ are obviously
all zero operators.

The Euler-Lagrange equations (36A,B) of paper~I  extracted from (16)
with $y_\mu = A_\mu (n)$ etc. yield:
$$
\eta^{\mu\nu} \Delta_{\mu}^{\scsc\#} \Delta_{\nu}^{\scsc\#} A_\sigma
(n) = 0\,,
\eqno{\rm (17A)}
$$
$$
\delta^{ab} \Delta_a^{\scsc\#} \Delta_b^{\scsc\#} A_\sigma
({\bfm{n}},t) - (\partial_t)^2 A_\sigma ({\bfm{n}},t) = 0\,.
\eqno{\rm (17B)}
$$
\vskip0.3cm

These equations are further augmented by the Lorentz-gauge constraint
on the allowable state vectors $|\psi \rangle$ (in a ``Hilbert space''
with indefinite metric):

$$
\langle \psi | \Delta_\mu^{\scsc\#} A^\mu (n) |\psi \rangle =
0\vspace*{0.2cm}\,,
\eqno{\rm (18A)}
$$
$$
\langle \psi | \Delta_b^{\scsc\#} A^b ({\bfm{n}},t) + \partial_t A^4
({\bfm{n}},t) |\psi \rangle = 0\,.
\eqno{\rm (18B)}
$$
\vskip0.3cm

\noindent
The Maxwell's equations (17A,B) and the Lorentz-gauge constraint
(18A,B) are preserved by the {\em restricted\/} gauge {\em
transformations\/} involving a hermitian operator $\Omega$:
$$
\widehat{A}_\mu (n) = A_\mu (n) - \Delta_\mu^{\scsc\#} \Omega
(n)\vspace*{0.2cm}\,,
\eqno{\rm (19A)}
$$
$$
\widehat{A}_j ({\bfm{n}},t) = A_j ({\bfm{n}},t) - \Delta_j^{\scsc\#}
\Omega ({\bfm{n}},t)\vspace*{0.2cm}\,,
\eqno{\rm (19Bi)}
$$
$$
\widehat{A}_4 ({\bfm{n}},t) = A_4 ({\bfm{n}},t) - \partial_t
\Omega ({\bfm{n}},t)\vspace*{0.3cm}\,,
\eqno{\rm (19Bii)}
$$
$$
\eta^{\mu\nu} \Delta_\mu^{\scsc\#} \Delta_\nu^{\scsc\#} \Omega (n)
= 0\vspace*{0.2cm}\,,
\eqno{\rm (20A)}
$$
$$
\delta^{ab} \Delta_a^{\scsc\#} \Delta_b^{\scsc\#} \Omega ({\bfm{n}},t)
- (\partial_t)^2 \Omega ({\bfm{n}},t) = 0\vspace*{0.2cm}\,,
\eqno{\rm (20B)}
$$
$$
\langle \psi | \Delta_\mu^{\scsc\#} \widehat{A}^\mu (n) | \psi \rangle
= 0\vspace*{0.2cm}\,,
\eqno(20\widehat{\mathrm A})
$$
$$
\langle \psi | \Delta_b^{\scsc\#} \widehat{A}^b ({\bfm{n}},t)
+ \partial_t \widehat{A}^4 ({\bfm{n}},t) | \psi \rangle = 0\,.
\eqno(20\widehat{\mathrm B})
$$
\vskip0.3cm

\noindent
By the Lagrangian (16) and the equations (39Biv) and (39Bvi) of
paper~I, we \vspace*{0.2cm}obtain
$$
P_j = \sum\limits_{{\bfm{n}}=0}^{\infty {(3)}}
\,\Big[(\Delta_j^{\scsc\#} A^\mu) \cdot (\partial_t
A_\mu)\Big]_{|t=0} \vspace*{0.2cm}\,,
\eqno{\rm (21Bi)}
$$
$$
H = -P_4 = (1/2) \sum\limits_{{\bfm{n}}=0}^{\infty {(3)}} \,\Big[
\delta^{ab} (\Delta_a^{\scsc\#} A^\mu)\cdot (\Delta_b^{\scsc\#}
A_\mu) + (\partial_t A^\mu)\cdot (\partial_t A_\mu )\Big]_{|t=0}
\,.
\eqno{\rm (21Bii)}
$$
\vskip0.3cm

\noindent
The above relativistic equations are {\em exact\/} and no additional
terms are necessary. The ``plane wave'' decomposition of the
four-potential operator $A_\mu ({\bfm{n}},t)$ is given by {[2]}:

$$
\begin{array}{l}
{\ds A_\mu ({\bfm{n}},t) = A_\mu^{\dagger} ({\bfm{n}},t)
= \int\limits_{\R^3} [2\nu ({\bfm{k}})]^{-1/2} \Biggr\{ a_\mu
({\bfm{k}}) \Biggr[\prod\limits_{\,j=1}^{3} \xi_{n^j} (k_j)\Biggr]
\!e^{-i\nu t} } \\[0.6cm]
{\ds + a_\mu^{\dagger} ({\bfm{k}}) \Biggr[
\prod\limits_{\,j=1}^{3} \,\overline{\xi_{n^j} (k_j)}\,\Biggr]
\!e^{i\nu t}\Biggr\} d^3 {\bfm{k}}
=: A_\mu^{-} ({\bfm{n}},t) + A_\mu^+ ({\bfm{n}},t)\,, } \\[0.6cm]
{\ds \nu ({\bfm{k}}) := + \sqrt{\delta^{ab} k_a k_b} =: -k_4 \,,}
\\[0.2cm]
{\ds \Delta_b^{\scsc\#} A_\mu ({\bfm{n}},t)_{|t=0} = i
\int\limits_{\R^3} [2\nu ({\bfm{k}})]^{1/2} k_b\Biggr\{a_\mu
({\bfm{k}}) \Biggr[ \prod\limits_{j} \xi_{n^j} (k_j)\Biggr] } \\[0.75cm]
{\ds - a_\mu^{\dagger} ({\bfm{k}}) \Biggr[ \prod\limits_{j}
\,\overline{\xi_{n^j}} (k_j)\Biggr] \Biggr\} d^3 {\bfm{k}}\,,
} \\[0.5cm]
\end{array}
$$
$$
\begin{array}{l}
{\ds \partial_t A_\mu ({\bfm{n}},t)_{|t=0} = i \int\limits_{\R^3}
{[\nu ({\bfm{k}}) / 2]^{1/2}} \Biggr\{a_\mu ({\bfm{k}}) \Biggr[
\prod\limits_{j} \xi_{n^j} (k_j)\Biggr] } \\[0.6cm]
{\ds -a_\mu^{\dagger} ({\bfm{k}}) \Biggr[ \prod\limits_{j}
\,\overline{\xi_{n^j}(k_j)}\,\Biggr] \Biggr\} d^3 {\bfm{k}}\,. }
\end{array}
\eqno\raisebox{-4ex}{\rm (22B)}
$$
\vskip0.4cm

The canonical quantization rules are assumed to be:
$$
\begin{array}{l}
{[a_\mu ({\bfm{k}}), a_\nu (\widehat{\bfm{k}})]} = [a_\mu^{\dagger}
({\bfm{k}}), a_\nu^{\dagger} (\widehat{\bfm{k}})] \equiv 0\,, \\[0.2cm]
{[a_\mu ({\bfm{k}}), a_\nu^{\dagger} (\widehat{\bfm{k}})]} =
\eta_{\mu\nu} \delta^3 ({\bfm{k}} - \widehat{\bfm{k}}) \,{\mathrm{I}}
\,({\bfm{k}})\,.
\end{array}
\eqno{\rm (23B)}
$$

The covariant commutation rules, which follow from (22B), (23B),
and (A.II.5B), are summarized below:
$$
\begin{array}{l}
\big[ A_\mu^+ ({\bfm{n}},t), A_\nu^+ (\widehat{\bfm{n}}, \widehat{t}\,)
\big] = \big[A_\mu^- ({\bfm{n}},t), A_\nu^- (\widehat{\bfm{n}},
\widehat{t}\,)\big] \equiv 0\,, \\[0.25cm]
\big[ A_\mu^- ({\bfm{n}},t), A_\nu^+ (\widehat{\bfm{n}}, \widehat{t}\,)
\big] = i \eta_{\mu\nu} D_+
({\bfm{n}},t;\,\widehat{\bfm{n}},\widehat{t}\,)\,{\mathrm{I}}\,,
\\[0.25cm]
\big[ A_\mu^+ ({\bfm{n}},t), A_\nu^- (\widehat{\bfm{n}}, \widehat{t}\,)
\big] = i \eta_{\mu\nu} D_-
({\bfm{n}},t;\,\widehat{\bfm{n}},\widehat{t}\,)\,{\mathrm{I}}\,,
\\[0.25cm]
\big[ A_\mu ({\bfm{n}},t), A_\nu (\widehat{\bfm{n}}, \widehat{t}\,)
\big] = i \eta_{\mu\nu} D
({\bfm{n}},t;\,\widehat{\bfm{n}},\widehat{t}\,)\,{\mathrm{I}}\,,
\\[0.25cm]
\big[ A_\mu ({\bfm{n}},t), A_\nu (\widehat{\bfm{n}}, \widehat{t}\,)
\big]_{|\hat{t}=t} \equiv 0  \qquad {\bfm{n}}\neq
\widehat{\bfm{n}}\,, \\[0.25cm]
\big[\partial_t A_\mu ({\bfm{n}},t), A_\nu (\widehat{\bfm{n}},
\widehat{t}\,) \big]_{|\hat{t}=t} = -i\eta_{\mu\nu} \delta_{{\bfm{n}}
\widehat{\bfm{n}}}^3\,{\mathrm{I}}\,, \\[0.25cm]
\big[\partial_t A_\mu ({\bfm{n}},t), \partial_{\hat{t}} A_\nu
(\widehat{\bfm{n}}, \hat{t}\,) \big]_{|\hat{t}=t} \equiv 0
\;\; \mbox{for} \;\; {\bfm{n}}\neq \widehat{\bfm{n}}\,.
\end{array}
\eqno{\rm (24B)}
$$

\noindent
The computations of the total momentum-energy from (21Bi,ii) and
(22B) yield:
$$
\begin{array}{l}
{\ds P_j = (1/2) \int\limits_{\R^3} \eta^{\mu\nu} \big[a_\mu^{\dagger}
({\bfm{k}}) a_\nu ({\bfm{k}}) + a_\mu ({\bfm{k}}) a_\nu^{\dagger}
({\bfm{k}})\big] \,k_j d^3 {\bfm{k}}\,,} \\[0.3cm]
{\ds H = -P_4 = (1/2) \int\limits_{\R^3} \eta^{\mu\nu}
\big[a_\mu^{\dagger} ({\bfm{k}}) a_\nu ({\bfm{k}}) + a_\mu ({\bfm{k}})
a_\nu^{\dagger} ({\bfm{k}})\big] \nu ({\bfm{k}})\,d^3 {\bfm{k}}\,.}
\end{array}
\eqno{\rm (25B)}
$$

\noindent
We can choose a special gauge so that (25B) simplifies considerably.
We assume the \vspace*{0.1cm}condition
$$
\langle \psi | k_\mu a^\mu (k) | \psi \rangle = 0\vspace*{0.1cm}\,.
\eqno(26)
$$
This is a {\em sufficient\/} condition for the satisfaction of the
Lorentz-gauge conditions (18A,B). We introduce a special restricted
gauge condition (see (22B)) with the help of the hermitian
operator-valued function
$$
\begin{array}{rcl}
\Omega ({\bfm{n}},t) &:=& {\ds -4i \int\limits_{\R^3} {[2\nu
({\bfm{k}})]}^{-5/2} k_b \Biggr\{ a^b ({\bfm{k}}) \Biggr[
\prod\limits_{j} \xi_{n^j} (k_j)\Biggr] e^{-i\nu t} }
\\[0.7cm]
&& {\ds - a^{\dagger b} ({\bfm{k}}) \Biggr[ \prod\limits_{j}
\,\overline{\xi_{n^j} (k_j)} \,\Biggr] e^{i\nu t} \Biggr\} d^3
{\bfm{k}}\,. }
\end{array}
\eqno{\rm (27B)}
$$

\noindent
Under this gauge transformation, the new field operators
$\widehat{a}_\mu ({\bfm{k}})$ (corresponding to the field
$\widehat{A}_\mu ({\bfm{n}},t)$)
undergo the following transformations:
$$
\widehat{a}_\mu ({\bfm{k}}) = a_\mu ({\bfm{k}}) - [\nu ({\bfm{k}})]^{-2}
k_\mu [k_b a^b ({\bfm{k}})]\,,
\eqno{\rm (28i)}
$$
$$
\langle \psi | k_\mu \widehat{a}^{\mu } ({\bfm{k}}) | \psi \rangle =
0\vspace*{0.15cm}\,,
\eqno{\rm (28ii)}
$$
$$
\langle \psi | \widehat{a}_4 ({\bfm{k}}) | \psi \rangle = 0\,.
\eqno{\rm (28iii)}
$$
\vskip0.2cm

\noindent
Thus, by (28iii) the temporal component $\widehat{a}_4 ({\bfm{k}})$
drops off. Next, let us consider the ``orthonormal'' tetrad {[8]}
$e_{(\lambda )}^\mu ({\bfm{k}})$ (also see equation (23) of paper~I)
which in general satisfy:
$$
\begin{array}{l}
\eta^{(\lambda\sigma)} e_{(\lambda )}^\mu ({\bfm{k}})
e_{(\sigma)}^{\nu} ({\bfm{k}}) = \eta^{\mu\nu}\,, \\[0.25cm]
\eta_{\mu\nu} e_{(\lambda)}^\mu ({\bfm{k}})
e_{(\sigma)}^{\nu} ({\bfm{k}}) = \eta_{(\lambda\sigma)}\,, \\[0.25cm]
a_{(\lambda)} ({\bfm{k}}) := a_{\mu} ({\bfm{k}}) e_{(\lambda)}^{\mu}
({\bfm{k}})\,, \\[0.25cm]
a^{\mu} ({\bfm{k}}) = e_{(\lambda)}^{\mu} ({\bfm{k}}) a^{(\lambda)}
({\bfm{k}})\,.
\end{array}
\eqno(29)
$$

We can choose prudently (for $\nu ({\bfm{k}})>0$) two of the tetrad
vectors by the following:
$$
\begin{array}{l}
\big(e_{(3)}^1 ({\bfm{k}}), \,e_{(3)}^2 ({\bfm{k}}), \,e_{(3)}^3
({\bfm{k}}), \,e_{(3)}^4 ({\bfm{k}})\big) := [\nu ({\bfm{k}})]^{-1}
(k_1, k_2, k_3, 0)\,, \\[0.25cm]
e_{(4)}^\mu ({\bfm{k}}) := \delta_{(4)}^\mu \,.
\end{array}
\eqno(30)
$$

\noindent
The choice of the other two vectors $e_{(1)}^\mu ({\bfm{k}})$ and
$e_{(2)}^\mu ({\bfm{k}})$ is arbitrary up to a two-dimensional
orthogonal transformation.

The condition (28ii), by (30) yields
$$
\begin{array}{rcl}
k_\mu e_{(\lambda)}^\mu ({\bfm{k}}) \langle \psi |
\widehat{a}^{(\lambda)} ({\bfm{k}}) | \psi \rangle &=& k_\mu e_{(3)}^\mu
({\bfm{k}}) \langle \psi | \widehat{a}^{(3)} ({\bfm{k}}) | \psi \rangle
\\[0.25cm]
&=& [\nu ({\bfm{k}})] \,\langle \psi | \widehat{a}^{(3)}({\bfm{k}})
| \psi \rangle = 0\,.
\end{array}
\eqno(31)
$$

\noindent
Thus, the expectation value of the longitudinal component $\langle
\psi | \widehat{a}^{(3)} ({\bfm{k}}) | \psi \rangle$ vanishes. Dropping
circumflexes in the sequence, we obtain from (25B), (28iii),
(31), and (29), the simplified versions of the expectation values
of the total momentum-energy as:
$$
\begin{array}{l}
{\ds \langle \psi | P_j | \psi \rangle = \sum\limits_{\lambda=1}^{2}
\;\int\limits_{\R^3} \langle \psi | N_{(\lambda )} ({\bfm{k}})|
\psi \rangle k_j d^3 {\bfm{k}}\,, } \\[0.6cm]
{\ds \langle \psi |H| \psi \rangle = \sum\limits_{\lambda=1}^{2}
\;\int\limits_{\R^3} \langle \psi | N_{(\lambda )} ({\bfm{k}})
+ (1/2)\,\delta^3 ({\bfm{0}})\, {\mathrm{I}}_{(\lambda )} ({\bfm{k}}|
\psi \rangle \nu ({\bfm{k}})\,d^3 {\bfm{k}}\,. }
\end{array}
\eqno{\rm (32B)}
$$

\noindent
Only the {\em two\/} degrees of (linear) polarization $(\lambda \in
\{1,2\})$ contribute to the total momentum-energy of the photons.
The equations in (32B) are identical to those obtained by the usual
relativistic theory {[7]}.
\vskip1.2cm

\section{Quantization of free spin-${\mathbf{1/2}}$ field}

The $4\times 4$ Dirac matrices $\gamma^\mu$ satisfy {[5]}:
$$
\begin{array}{l}
\gamma^\mu \gamma^\nu + \gamma^\nu \gamma^\mu = 2\eta^{\mu\nu}
\,{\mathrm{I}}\,, \\[0.2cm]
{[\gamma^j]}^{\dagger} = \gamma^j, \, [\gamma^4]^{\dagger}
= -\gamma^4\,.
\end{array}
\eqno(33)
$$

\noindent
The Dirac bispinor field $\rho =\psi (n)$ or $\rho = \psi
({\bfm{n}},t)$ is a $4\times 1$ column vector of {\em operators\/}
in the second quantized theory. We denote
$$
\begin{array}{l}
\tilde{\rho} := i\rho^{\dagger} \gamma^4\,, \\[0.2cm]
{[\tilde{\rho} \rho]}^{\dagger} = \tilde{\rho} \rho\,.
\end{array}
\eqno(34)
$$

The Lagrangian function for a massive, spin-1/2 field operator
$\rho$ is taken to be:
$$
\begin{array}{l}
L(\rho , \tilde{\rho}; \,\rho_\mu, \tilde{\rho}_\mu) :=
-(1/2) (\tilde{\rho} \gamma^\mu \rho_\mu - \tilde{\rho}_\mu \gamma^\mu
\rho) - m\tilde{\rho} \rho ; \;\; m>0\,, \\[0.4cm]
{\ds \frac{\partial L(..)}{\partial \rho} = (1/2)
\tilde{\rho}_\mu \gamma^\mu - m\tilde{\rho}, \;\; \frac{\partial
L(..)}{\partial \tilde{\rho}} = -(1/2) \gamma^\mu \rho_\mu
- m\rho \,, } \\[0.6cm]
{\ds \frac{\partial L(..)}{\partial \rho_\mu} } =
{\ds -(1/2)\tilde{\rho} \gamma^\mu, \;\; \frac{\partial L(..)}{\partial
\tilde{\rho}_\mu} = (1/2) \gamma^\mu \rho\,, } \\[0.6cm]
{\ds \frac{\partial^2 L(..)}{\partial\tilde{\rho} \partial \rho}}
= {\ds - m\,{\mathrm{I}} = \frac{\partial^2 L(..)}{\partial
\rho \partial \tilde{\rho}}, \;\; \frac{\partial^2 L(..)}{\partial
\tilde{\rho} \partial \rho_\mu} = \frac{\partial^2 L(..)}{\partial
\rho_\mu \partial \tilde{\rho}} = -(1/2) \gamma^\mu, } \\[0.65cm]
{\ds \frac{\partial^2 L(..)}{\partial
\tilde{\rho}_\mu \partial \rho} = \frac{\partial^2 L(..)}{\partial
\rho \partial \tilde{\rho}_\mu} = (1/2) \gamma^\mu\,, } \\[0.65cm]
{\ds \frac{\partial^2 L(..)}{\partial \rho_\mu \partial
\rho_\nu} } = {\ds  \frac{\partial^2 L(..)}{\partial
\tilde{\rho}_\mu \partial \tilde{\rho}_\nu} = \frac{\partial^2
L(..)}{\partial \tilde{\rho}_{\mu} \partial \rho_{\nu}}
= \frac{\partial^2 L(..)}{\partial \rho_\mu \partial \tilde{\rho}_\nu}
\equiv 0\,. }
\end{array}
\eqno(35)
$$
\vskip0.3cm

\noindent
The triple and higher partial derivatives of $L$ are all identically
zero operators.

The Euler-Lagrange operator equations from (35) and equations
(38A,$\overline{\rm A}$), and (38B,$\overline{\rm B}$) of paper~I
are furnished by:
$$
\gamma^\mu  \Delta_\mu^{\scsc\#} \psi (n) + m\psi (n) = 0\,,
\eqno{\rm (36A)}
$$
$$
{[\Delta_\mu^{\scsc\#} \tilde{\psi} (n)]}\,\gamma^\mu
- m \tilde{\psi} (n) = 0\vspace*{0.2cm}\,,
\eqno(36\overline{\rm A})
$$
$$
\gamma^j \Delta_j^{\scsc\#} \psi ({\bfm{n}},t) + \gamma^4 \partial_t
\psi ({\bfm{n}},t) + m\psi ({\bfm{n}},t) = 0\vspace*{0.2cm}\,,
\eqno{\rm (36B)}
$$
$$
{[\Delta_j^{\scsc\#} \tilde{\psi} ({\bfm{n}},t)]}\,\gamma^j
+ {[\partial_t \tilde{\psi } ({\bfm{n}},t)]}\,\gamma^4
- m\tilde{\psi} ({\bfm{n}},t) = 0\vspace*{0.3cm}\,.
\eqno(36\overline{\rm B})
$$

By the equation (35), and equations (A.II.5A) and (A.II.6Bi,ii) of
paper~I, we derive:

$$
\hspace*{-0.2cm}\begin{array}{l}
{\ds \Biggr\{ \Delta_\nu^{\scsc\#} \Biggr[\frac{\partial L(..)}{\partial
\rho_\nu}\Biggr]_{|..}\!\cdot \Delta_\mu^{\scsc\#} \psi + \Biggr[
\frac{\partial L(..)}{\partial \rho_\nu}\Biggr]_{|..}\!\cdot
\Delta_\mu^{\scsc\#} \Delta_\nu^{\scsc\#} \psi + {\mathrm{(h.c.)}}
- \Delta_\mu^{\scsc\#} L(..)_{|..}\Biggr\} }\\[0.6cm]
{\ds + \frac{\partial^2 L(..)}{\partial \tilde{\rho} \partial \rho_{|..}}
{\,[\Delta_\mu^{\scsc\#} (\tilde{\psi}(n) \cdot \psi (n)) - \tilde{\psi}
(n) \Delta_\mu^{\scsc\#} \psi - (\Delta_\mu^{\scsc\#} \tilde{\psi})
\cdot \psi (n)]} } \\[0.6cm]
{\ds + \Biggr\{ \Delta_\mu^{\scsc\#} \Biggr[\tilde{\psi}(n)
\cdot\Biggr( \frac{\partial^2 L(..)}{\partial \tilde{\rho} \partial
\rho_\nu}\Biggr)_{\!\!|..} \cdot \Delta_\nu^{\scsc\#} \psi \Biggr]
} \\[0.6cm]
{\ds - \tilde{\psi}(n) \cdot \Delta_\mu^{\scsc\#} \Biggr[\Biggr(
\frac{\partial^2 L(..)}{\partial \tilde{\rho} \partial
\rho_\nu}\Biggr)_{\!\!|..} \Delta_\nu^{\scsc\#} \psi \Biggr]
- \Delta_\mu^{\scsc\#} \tilde{\psi} \cdot \Biggr[
\frac{\partial^2 L(..)}{\partial \tilde{\rho} \partial
\rho_\nu}\Biggr]_{|..} \Delta_\nu^{\scsc\#}\psi \Biggr\} }\\[0.6cm]
{\ds + \Biggr\{ \Delta_\mu^{\scsc\#} \Biggr[\Delta_\nu^{\scsc\#}
\tilde{\psi} \Biggr( \frac{\partial^2 L(..)}{\partial \tilde{\rho}_\nu
\partial \rho}\Biggr)_{\!\!|..} \psi (n)\Biggr] } \\[0.6cm]
{\ds - \Delta_\nu^{\scsc\#} \tilde{\psi}\cdot \Biggr[
\frac{\partial^2 L(..)}{\partial \tilde{\rho}_\nu \partial
\rho}\Biggr]_{|..} \cdot \Delta_\mu^{\scsc\#} \psi
- (\Delta_\mu^{\scsc\#} \Delta_\nu^{\scsc\#} \tilde{\psi}) \cdot
\Biggr[ \frac{\partial^2 L(..)}{\partial \tilde{\rho}_\nu \partial
\rho}\Biggr]_{|..} \cdot \psi (n) \Biggr\} }\\[0.6cm]
+ 0 =: \Delta_\nu T_\mu^\nu  (n) = 0\,,
\end{array}
\eqno{\rm (37A)}
$$
\vskip0.5cm

$$
\Delta_b T_a^b ({\bfm{n}},t) + \partial_t T_a^4 ({\bfm{n}},t) =
0\vspace*{0.3cm}\,,
\eqno{\rm (37Bi)}
$$
$$
\Delta_b T_4^b ({\bfm{n}},t) + \partial_t T_4^4 ({\bfm{n}},t) = 0\,,
\eqno{\rm (37Bii)}
$$
\vskip0.5cm

$$
\begin{array}{l}
{\ds T_a^b ({\bfm{n}},t) := \sqrt{\frac{n^b}{2}} \Biggr\{
\Biggr[\frac{\partial L(..)}{\partial \rho_{b|(..,n^b-1,..)}} \cdot
\Delta_a^{\scsc\#} \psi } \\[0.6cm]
{\ds + \frac{\partial L(..)}{\partial
\rho_{b|..}} \cdot (\Delta_a^{\scsc\#}\psi )_{|(..,n^b-1,..)}
+ {\mathrm{(h.c.)}}\Biggr] } \\[0.6cm]
{\ds -\delta_a^b \Biggr[ \frac{1}{2} (\tilde{\psi} (..,n^b-1,..)\cdot
\gamma^c \Delta_c^{\scsc\#} \psi + \tilde{\psi} ({\bfm{n}},t) \cdot
\gamma^c (\Delta_c^{\scsc\#} \psi )_{|(..,n^b-1,..)} }
\\[0.6cm]
+ \tilde{\psi} (..,n^b-1,..) \gamma^4 \partial_t \psi  + \tilde{\psi}
({\bfm{n}},t) \gamma^4 (\partial_t \psi )_{|(..,n^b-1,..)}  \\[0.4cm]
- (\Delta_c^{\scsc\#} \tilde{\psi})_{|(..,n^b-1,..)} \cdot \gamma^c
\psi ({\bfm{n}},t) \\[0.5cm]
- (\Delta_c^{\scsc\#} \tilde{\psi}) \gamma^c \psi (..,n^b-1,..)
- (\partial_t \tilde{\psi})_{|(..,n^b-1,..)} \cdot \gamma^4 \cdot
\psi ({\bfm{n}},t) \\[0.5cm]
- (\partial_t \tilde{\psi}) \cdot \gamma^4 \cdot \psi
(..,n^b-1,..) \\[0.25cm]
+ m(\tilde{\psi}(..,n^b-1,..) \cdot \psi ({\bfm{n}},t)
+ \tilde{\psi} ({\bfm{n}},t) \cdot \psi (..,n^b-1,..)\Biggr]\Biggr\},
\end{array}
\eqno{\rm (37Biii)}
$$
\vskip0.5cm

$$
T_a^4 ({\bfm{n}},t) := \frac{\partial L(..)}{\partial \rho_{4|..}}
\cdot \Delta_a^{\scsc\#} \psi  + (\Delta_a^{\scsc\#} \tilde{\psi})
\cdot \frac{\partial L(..)}{\partial
\tilde{\rho}_{4|..}}\vspace*{0.2cm}\,,
\eqno{\rm (37Biv)}
$$

$$
\begin{array}{l}
{\ds T_4^b ({\bfm{n}},t) := \sqrt{\frac{n^b}{2}} \Biggr[\frac{\partial
L(..)}{\partial \rho_{b|(..,n^b-1,..)}} \cdot \partial_t \psi} \\[0.3cm]
{\ds + \frac{\partial L(..)}{\partial \rho_{b|..}} \cdot
(\partial_t \psi)_{|(..,n^b-1,..)} + {\mathrm (h.c.)}\Biggr],}
\end{array}
\eqno{\rm (37Bv)}
$$

$$
T_4^4 ({\bfm{n}},t) := \left[ \frac{\partial L(..)}{\partial
\rho_{4|..}} \cdot (\partial_t \psi)
+ (\partial_t \tilde{\psi}) \cdot \frac{\partial L(..)}{\partial
\tilde{\rho}_{4|..}} - L(..)_{|..}\right].
\eqno{\rm (37Bvi)}
$$
\vskip0.2cm

\noindent
The above equations are {\em exact}. Note that the {\em tricky\/}
ordering of the operators in the above equations is {\em crucial}.

The {\em relativistic\/} total momentum-energy and the charge are
given by (see equations (7Bi,ii,iii) and (37Biv,vi)):
$$
P_j = (i/2) \sum\limits_{{\bfm{n}}=0}^{\infty (3)} \,\big[
\Delta_j^{\scsc\#} \psi^{\dagger} \cdot \psi ({\bfm{n}},t)
- \psi^{\dagger} ({\bfm{n}},t) \cdot \Delta_j^{\scsc\#}
\psi \big]_{|t=0} \,,
\eqno{\rm (38Bi)}
$$
$$
H = -P_4 = (i/2) \sum\limits_{{\bfm{n}}=0}^{\infty (3)} \,\big[
\psi^{\dagger} ({\bfm{n}},t) \cdot \partial_t \psi - \partial_t
\psi^{\dagger} \cdot \psi ({\bfm{n}},t) \big]_{|t=0} \,,
\eqno{\rm (38Bii)}
$$
$$
Q = e \sum\limits_{{\bfm{n}}=0}^{\infty (3)} \psi^{\dagger}
({\bfm{n}},0) \cdot \psi ({\bfm{n}},0)\,.
\eqno{\rm (38Biii)}
$$
\vskip0.3cm

A class of ``plane wave'' solutions of the Dirac equations (36A)
and (36B) is provided by (see reference [2] and Appendix~I):
$$
\begin{array}{l}
{\ds \psi (n) = \int\limits_{\R^3} [m / E({\bfm{p}})]^{1/2} \Biggr\{
\sum\limits_{r=1}^{2} \,\Biggr[ \alpha_r ({\bfm{p}}) {\bfm{u}}_r
({\bfm{p}}) \Biggr(\prod\limits_{\mu} \xi_{n^\mu} (p_\mu)\Biggr)
} \\[0.6cm]
{\ds + \beta_r^{\dagger} ({\bfm{p}}) {\bfm{v}}_r
({\bfm{p}}) \Biggr(\prod\limits_{\mu} \,\overline{\xi_{n^\mu}
(p_\mu)}\Biggr) \Biggr]\Biggr\}d^3 {\bfm{p}} } \\[0.6cm]
=: \psi^{-} (n) + \psi^{+} (n)\,,
\end{array}
\eqno{\rm (39Ai)}
$$

$$
\begin{array}{l}
{\ds \tilde{\psi} (n) = \int\limits_{\R^3} [m / E({\bfm{p}})]^{1/2}
\Biggr\{ \sum\limits_{r=1}^{2} \,\Biggr[ \alpha_r^{\dagger} ({\bfm{p}})
\tilde{\bfm{u}}_r ({\bfm{p}}) \Biggr(\prod\limits_{\mu}
\overline{\xi_{n^\mu} (p_\mu)}\Biggr) } \\[0.6cm]
{\ds + \beta_r ({\bfm{p}}) \tilde{\bfm{v}}_r
({\bfm{p}}) \Biggr(\prod\limits_{\mu} \xi_{n^\mu}
(p_\mu)\Biggr) \Biggr]\Biggr\}d^3 {\bfm{p}} } \\[0.6cm]
=: \tilde{\psi}^{+} (n) + \tilde{\psi}^{-} (n)\,,
\end{array}
\eqno{\rm (39Aii)}
$$

$$
\begin{array}{l}
{\ds \psi ({\bfm{n}},t) := \int\limits_{\R^3} [m / E({\bfm{p}})]^{1/2}
\Biggr\{ \sum\limits_{r=1}^{2} \,\Biggr[\alpha_r ({\bfm{p}}) {\bfm{u}}_r
({\bfm{p}}) \Biggr(\prod\limits_{j} \xi_{n^j} (p_j)\Biggr) e^{-iEt}
} \\[0.6cm]
{\ds + \beta_r^{\dagger} ({\bfm{p}}) {\bfm{v}}_r
({\bfm{p}}) \Biggr(\prod\limits_{j} \,\overline{\xi_{n^j}
(p_j)}\Biggr) e^{iEt} \Biggr]\Biggr\} d^3 {\bfm{p}} } \\[0.6cm]
=: \psi^{-} ({\bfm{n}},t) + \psi^{+} ({\bfm{n}},t)\,,
\end{array}
\eqno{\rm (39Bi)}
$$
\vskip0.2cm

$$
\begin{array}{l}
{\ds \tilde{\psi} ({\bfm{n}},t) = \int\limits_{\R^3}
{[m / E({\bfm{p}})]}^{1/2} \Biggr\{ \sum\limits_{r=1}^{2}
\,\Biggr[ \alpha_r^{\dagger} ({\bfm{p}}) \tilde{\bfm{u}}_r
({\bfm{p}}) \Biggr(\prod\limits_{j} \overline{\xi_{n^j} (p_j)}\Biggr)
e^{iEt} } \\[0.6cm]
{\ds + \beta_r ({\bfm{p}}) \tilde{\bfm{v}}_r
({\bfm{p}}) \Biggr(\prod\limits_{j} \,\xi_{n^j} (p_j)\Biggr) e^{-iEt}
\Biggr]\Biggr\} d^3 {\bfm{p}} } \\[0.6cm]
=: \tilde{\psi}^{+} ({\bfm{n}},t) + \tilde{\psi}^{-} ({\bfm{n}},t)\,,
\end{array}
\eqno{\rm (39Bii)}
$$
\vskip0.4cm

$$
\begin{array}{l}
{\ds p^4 = -p_4 \equiv E({\bfm{p}}) := + \sqrt{\delta^{ab} p_a p_b
+ m^2} > 0\,, } \\[0.25cm]
{\bfm{u}}_r^{\dagger} ({\bfm{p}}) {\bfm{u}}_s ({\bfm{p}})
= {\bfm{v}}_r^{\dagger} ({\bfm{p}}) {\bfm{v}}_s ({\bfm{p}})
= [E ({\bfm{p}}) / m]\,\delta_{rs} \:, \\[0.25cm]
{\bfm{u}}_r^{\dagger} ({\bfm{p}}) {\bfm{v}}_s (-{\bfm{p}})
= {\bfm{v}}_r^{\dagger} ({\bfm{p}}) {\bfm{u}}_s (-{\bfm{p}})
\equiv 0\,.
\end{array}
\eqno{\rm (39Biii)}
$$
\vskip0.5cm

The canonical quantization rules for a spin-1/2 field operator are
furnished by the anti-commutators:
$$
\begin{array}{rcl}
{[A,B]}_{+} &:=& AB + BA\,, \\[0.25cm]
{[\alpha_r ({\bfm{p}}), \alpha_s (\widehat{\bfm{p}})]}_{+}
&=& [\beta_r ({\bfm{p}}), \beta_s (\widehat{\bfm{p}})]_{+} \\[0.25cm]
&=& [\alpha_r^{\dagger} ({\bfm{p}}), \alpha_s^{\dagger}
(\widehat{\bfm{p}})]_{+} = [\beta _r^{\dagger} ({\bfm{p}}),
\beta_s^{\dagger} (\widehat{\bfm{p}})]_{+} \equiv 0\,, \\[0.25cm]
{[\alpha_r ({\bfm{p}}), \beta_s (\widehat{\bfm{p}})]}_{+}
&=& [\alpha_r^{\dagger} ({\bfm{p}}), \beta_s^{\dagger}
(\widehat{\bfm{p}})]_{+} \\[0.25cm]
&=& [\alpha_r ({\bfm{p}}), \beta_s^{\dagger}
(\widehat{\bfm{p}})]_{+} = [\alpha_r^{\dagger} ({\bfm{p}}),
\beta_s (\widehat{\bfm{p}})]_{+} \equiv 0\,, \\[0.25cm]
{[\alpha_r ({\bfm{p}}), \alpha_s^{\dagger} (\widehat{\bfm{p}})]}_{+}
&=& [\beta_r ({\bfm{p}}), \beta_s^{\dagger} (\widehat{\bfm{p}})]_{+}
= \delta_{rs} \delta^3 ({\bfm{p}} - \widehat{\bfm{p}}) \,{\mathrm{I}}\,.
\end{array}
\eqno(40)
$$
\vskip0.5cm

The particle and anti-particle occupation number operators are defined
by:
$$
\begin{array}{rcl}
N_r^{-} ({\bfm{p}}) &:=& \alpha_r^{\dagger} ({\bfm{p}}) \alpha_r
({\bfm{p}})\,, \\[0.25cm]
N_r^{+} ({\bfm{p}}) &:=& \beta_r^{\dagger} ({\bfm{p}}) \beta_r
({\bfm{p}})\,.
\end{array}
\eqno(41)
$$
\vskip0.2cm

\noindent
Here, the subscript ``$r$'' is {\em not\/} summed. The occupation
number operators $N_r^{-} ({\bfm{p}})$ and $N_r^{+} ({\bfm{p}})$ take
eigenvalues from $\{0,1\}.$ Therefore, this quantization is compatible
with the Fermi-Dirac statistics. The particle and anti-particle vacuum
state $| \psi_0\rangle$ is characterized by:
$$
\begin{array}{l}
\alpha _r ({\bfm{p}}) | \psi_0 \rangle = \beta_r ({\bfm{p}})
|\psi_0 \rangle = | {\bfm{0}} \rangle\,,  \\[0.25cm]
\langle \psi_0 | \psi_0\rangle = 1, \quad \langle {\bfm{0}}|{\bfm{0}}
\rangle = 0\,.
\end{array}
\eqno(42)
$$
\vskip0.2cm

We can derive the covariant quantization rules for a spin-1/2 particle
field by equations (40), (39A), and (39Bi,ii). These are provided
by the following anti-commutation relations:
$$
{[\psi^{-} (n), \tilde{\psi}^{-} (\widehat{n})]}_{+} = [\psi^{+} (n),
\tilde{\psi}^{+} (\widehat{n})]_{+} \equiv 0\vspace*{0.1cm}\,,
\eqno{\rm (43Ai)}
$$
$$
{[\psi^{-} (n), \psi^{+} (\widehat{n})]}_{+} = [\tilde{\psi}^{-} (n),
\tilde{\psi}^{+} (\widehat{n})]_{+} \equiv 0\vspace*{0.2cm}\,,
\eqno{\rm (43Aii)}
$$
$$
{[\psi (n), \psi (\widehat{n})]}_{+} = [\tilde{\psi} (n),
\tilde{\psi} (\widehat{n})]_{+} \equiv 0\vspace*{0.2cm}\,,
\eqno{\rm (43Aiii)}
$$
$$
{[\psi^{-} (n), \tilde{\psi}^{+} (\widehat{n})]}_{+} = iS_{+}
(n,\widehat{n}) \,{\mathrm{I}}\vspace*{0.2cm}\,,
\eqno{\rm (43Aiv)}
$$
$$
{[\psi^{+} (n), \tilde{\psi}^{-} (\widehat{n})]}_{+} = -iS_{-}
(n,\widehat{n}) \,{\mathrm{I}}\vspace*{0.2cm}\,,
\eqno{\rm (43Av)}
$$
$$
{[\psi (n), \tilde{\psi} (\widehat{n})]}_{+} = -iS
(n,\widehat{n}) \,{\mathrm{I}}\vspace*{0.2cm}\,,
\eqno{\rm (43Avi)}
$$
$$
{[\psi^{-} ({\bfm{n}},t), \tilde{\psi}^{-} (\widehat{\bfm{n}},t\,)]}_{+}
= [\psi^{+} ({\bfm{n}},t), \tilde{\psi}^{+} (\widehat{\bfm{n}},
\widehat{t})]_{+} \equiv 0\vspace*{0.2cm}\,,
\eqno{\rm (43Bi)}
$$
$$
{[\psi^{-} ({\bfm{n}},t), \psi^{+} ({\bfm{n}},\widehat{t}\,)]}_{+}
= [\tilde{\psi}^{-} ({\bfm{n}},t), \tilde{\psi}^{+} (\widehat{\bfm{n}},
\widehat{t}\,)]_{+} \equiv 0\vspace*{0.2cm}\,,
\eqno{\rm (43Bii)}
$$
$$
{[\psi ({\bfm{n}},t), \psi (\widehat{\bfm{n}},\widehat{t}\,)]}_{+}
= [\tilde{\psi} ({\bfm{n}},t), \tilde{\psi} (\widehat{\bfm{n}},
\widehat{t}\,)]_{+} \equiv 0\vspace*{0.2cm}\,,
\eqno{\rm (43Biii)}
$$
$$
{[\psi^{-} ({\bfm{n}},t), \tilde{\psi}^{+} (\widehat{\bfm{n}},
\widehat{t}\,)]}_{+} = -iS_{+} ({\bfm{n}},t; \widehat{\bfm{n}},
\widehat{t}\,) \,{\mathrm{I}}\vspace*{0.2cm}\,,
\eqno{\rm (43Biv)}
$$
$$
{[\psi^{+} ({\bfm{n}},t), \tilde{\psi}^{-} (\widehat{\bfm{n}},
\widehat{t}\,)]}_{+} = -iS_{-}(..) ({\bfm{n}},t; \widehat{\bfm{n}},
\widehat{t}\,) \,{\mathrm{I}}\vspace*{0.2cm}\,,
\eqno{\rm (43Bv)}
$$
$$
{[\psi ({\bfm{n}},t), \tilde{\psi} (\widehat{\bfm{n}},
\widehat{t}\,)]}_{+} = -iS ({\bfm{n}},t; \widehat{\bfm{n}},
\widehat{t}\,) \,{\mathrm{I}}\vspace*{0.2cm}\,,
\eqno{\rm (43Bvi)}
$$
$$
{[\psi ({\bfm{n}},0), \psi^{\dagger} (\widehat{\bfm{n}},0)]}_{+}
= \delta_{{\bfm{n}} \hat{\bfm{n}}}^3 \,{\mathrm{I}}\,.
\eqno{\rm (43Bvii)}
$$
\vskip0.3cm

\noindent
Here, the Green's functions $\mathop{S}\limits_{(a)} (n,\widehat{n})$
and $\mathop{S}\limits_{(a)} ({\bfm{n}},t; \widehat{\bfm{n}},
\widehat{t}\,)$ are \vspace*{-0.2cm}borrowed from Appendix~II.

Now, we shall compute the total momentum, energy, and charge by the
equations (39Bi), (38Bi,ii,iii), (40), and (41). After laborious
calculations, similar to those in Appendix~III, we obtain rather neat
results:
$$
P_j = \sum\limits_{r=1}^{2} \;\int\limits_{\R^3} \,[N_r^{-}
({\bfm{p}}) + N_r^{+} ({\bfm{p}})] \,p_j d^3 {\bfm{p}}\,,
\eqno{\rm (44Bi)}
$$
$$
H = -P_4 = \sum\limits_{r=1}^{2} \;\int\limits_{\R^3} \,[N_r^{-}
({\bfm{p}}) + N_r^{+} ({\bfm{p}}) - \delta^3 ({\bfm{0}})
\,{\mathrm{I}}_r ({\bfm{p}})] \,E ({\bfm{p}}) d^3 {\bfm{p}}\,,
\eqno{\rm (44Bii)}
$$
$$
Q = e \sum\limits_{r=1}^{2} \;\int\limits_{\R^3} \,[N_r^{-}
({\bfm{p}}) - N_r^{+} ({\bfm{p}}) + \delta^3 ({\bfm{0}})
\,{\mathrm{I}}_r ({\bfm{p}})] \,d^3 {\bfm{p}}\,,
\eqno{\rm (44Biii)}
$$

\noindent
In the case of the electron-position field, we choose $N_r^{-}
({\bfm{p}})$ as the number operator for electrons, and the charge
parameter $e = -\sqrt{4\pi / 137}$; in contrast to the charged scalar
particles in (13iii). The results (44Bi,ii,iii) {\em coincide
exactly\/} with the usual relativistic quantum theory of a free
spin-1/2 field in the (flat) space-time continuum {[7]}.
\vskip1.2cm

\section*{Appendix I: \ Hermite and related complex polynomials}

The definitions and the useful formulae for the Hermite polynomials
{[9]} are provided here $(k\in \R; \; n\in \{0,1,2,\ldots\})$:
$$
H_n (k) := (-1)^n e^{k^2} \frac{d^n}{(dk)^n} \,(e^{-k^2})\,,
\eqno{\rm (A.I.1)}
$$
$$
H_n (-k) := (-1)^n H_n (k)\vspace*{0.3cm}\,,
\eqno{\rm (A.I.2)}
$$
$$
\frac{d^2 H_n (k)}{d k^2} - 2k \frac{d H_n (k)}{d k} + 2n H_n (k)
= 0\vspace*{0.3cm}\,,
\eqno{\rm (A.I.3)}
$$
$$
\frac{d H_n (k)}{d k} = 2n H_{n-1} (k), \quad n\geq 1\vspace*{0.3cm}\,,
\eqno{\rm (A.I.4)}
$$
$$
H_{n+1} (k) = 2k H_n (k) - 2n H_{n-1} (k), \quad n\geq 1\vspace*{0.2cm}\,.
\eqno{\rm (A.I.5)}
$$

\noindent
The complex, orthonormal polynomials $\xi_n (k)$ satisfy the following
definition and equations:
$$
\xi_n(k)\!:=\!\frac{(i)^n e^{-k^2/2} H_n(k)}{\pi^{1/4}\,2^{n/2}
\sqrt{n!}}\!=:\!(i)^n f_n(k), \;\; \xi_{-n-1}(k) := 0, \;\;\mbox{for}
\;\; n\in \N,\;
\eqno{\rm (A.I.6)}
$$
$$
\xi_n (-k) = (-1)^n \xi_n (k) = \overline{\xi_n (k)}\vspace*{0.2cm}\,,
\eqno{\rm (A.I.7)}
$$
$$
\frac{d^2 \xi_n (k)}{dk^2} + (-k^2 + 2n + 1) \xi_n (k)
= 0\vspace*{0.2cm}\,,
\eqno{\rm (A.I.8)}
$$
$$
\frac{d \xi_n (k)}{dk} + k \xi_n (k) = i \sqrt{2n} \:\xi_{n-1}
(k)\vspace*{0.2cm}\,,
\eqno{\rm (A.I.9)}
$$
$$
\Delta^{\scsc\#} \xi_n (k) = ik \xi_n (k)\vspace*{0.5cm}\,,
\eqno{\rm (A.I.10)}
$$
$$
\begin{array}{l}
\xi_{2n} (k) = {\ds \frac{(-1)^n e^{-k^2/2}}{\pi^{1/4}\,2^n
\sqrt{2n!}} \: \cdot } \\ [0.4cm]
{\ds \cdot \: \Bigg[(2k)^{2n} + \sum\limits_{j=1}^{n} (-1)^j
2^{2n-j} (2j-1) !! \left({2n \atop 2j}\right) k^{2(n-j)}
\Bigg],}
\end{array}
\eqno{\rm (A.I.11a)}
$$
\vskip0.2cm

$$
\begin{array}{l}
\xi_{2n+1} (k) = {\ds \frac{i(-1)^n e^{-k^2/2}}{\pi^{1/4}
\,2^{n+1/2} \sqrt{(2n+1)!}} \: \cdot } \\[0.4cm]
{\ds \cdot \: \Bigg[(2k)^{2n+1}+\sum\limits_{j=1}^{n} (-1)^j
2^{2n+1-j} (2j-1) !! \left({2n+1 \atop 2j}\right)
k^{2n+1-2j} \Bigg],} \\[0.6cm]
((2j-1)!! := (2j-1)(2j-3) \ldots 5.3.1\;.)
\end{array}
\eqno{\rm (A.I.11b)}
$$
\vskip0.5cm

$$
\xi_0 (k) = \frac{e^{-k^2/2}}{\pi^{1/4}}\vspace*{0.2cm} \,,
\eqno{\rm (A.I.12)}
$$
$$
\xi_1 (k) = \frac{i \sqrt{2}\,k \,e^{-k^2/2}}{\pi^{1/4}} \vspace*{0.2cm}\,,
\eqno{\rm (A.I.13)}
$$
$$
\xi_{2n} (0) = \frac{(2n-1)!!}{\pi^{1/4} \,\sqrt{(2n)!}}
= \frac{\sqrt{(2n)!}}{\pi^{1/4}\, 2^nn!} \vspace*{0.3cm}\,,
\eqno{\rm (A.I.14a)}
$$
$$
\xi_{2n+1} (0) \equiv 0\vspace*{0.3cm}\,,
\eqno{\rm (A.I.14b)}
$$
$$
\exp \big\{[(t^2+k^2)/2] + i \sqrt{2} \,tk\big\}
= \sum\limits_{n=0}^{\infty} \,\frac{\xi_n(k) t^n}{\sqrt{n!}}
\vspace*{0.4cm}\,,
\eqno{\rm (A.I.15)}
$$
$$
\begin{array}{l}
{\ds 2^{n/2} \exp \big\{-[(k-p)/2]^2\big\} \,\xi_n [(k+p)/
\sqrt{2}\,] } \\[0.2cm]
{\ds = \pi^{1/4} \sum\limits_{j=0}^{n} \,\sqrt{\left({n \atop j}
\right)} \,\xi_{n-j} (k) \xi_j (p)\,, }
\end{array} \vspace*{0.25cm}\,
\eqno{\rm (A.I.16)}
$$

$$
\begin{array}{l}
{\ds (k-\widehat{k}) \sum\limits_{n=0}^{N} \,\overline{\xi_n (k)}
\, \xi_n (\widehat{k}) = i \sqrt{(N+1)/2} } \\[0.3cm]
{\ds \left[\,\overline{\xi_{N+1}(k)} \,\xi_N (\widehat{k})
- \overline{\xi_{N+1}(\widehat{k})} \,\xi_N (k) \right],}
\end{array}
\eqno{\rm (A.I.17)}
$$
\vskip0.3cm

$$
\int\limits_{-\infty}^{\infty} \overline{\xi_m(k)} \,\xi_n(k)\,dk
= \int\limits_{-\infty}^{\infty} f_m(k) f_n(k)\,dk =
\delta_{mn}\vspace*{0.2cm}\,,
\eqno{\rm (A.I.18)}
$$
$$
\sum\limits_{n=0}^{\infty} \overline{\xi_n(k)} \,\xi_n(p) = \delta
(k-p)\vspace*{0.2cm}\,,
\eqno{\rm (A.I.19)}
$$
$$
\sum\limits_{n=0}^{\infty} \xi_n(k) \xi_n(p)
= \sum\limits_{n=0}^{\infty} \overline{\xi_n(k)} \,
\, \overline{\xi_n(p)} = \delta (k+p)\vspace*{0.2cm}\,,
\eqno{\rm (A.I.20)}
$$
$$
\xi_n(k) = \frac{1}{\sqrt{2\pi}} \, \int\limits_{-\infty}^{\infty}
f_n(x) \,e^{ikx} dx\,.
\eqno{\rm (A.I.21)}
$$
\vskip1.2cm

\section*{Appendix II: \ Non-singular Green's functions}

The relativistic invariant Green's functions [2] for the finite
difference and the difference-differential Klein-Gordon equations
(4A) and (4B) respectively are given by (see Fig.\,1):
$$
\begin{array}{l}
{\ds \mathop{\Delta}\limits_{(a)} (n; \widehat{n}; \mu ) :=
\int\limits_{\R^3} \Biggr\{\, \int\limits_{C_{(a)}}
(\eta^{\alpha\beta} k_\alpha k_\beta + \mu^2)^{-1} } \\[0.6cm]
{\ds \Biggr[\,\prod\limits_{\mu=1}^{4} \xi_{n^\mu}(k_\mu)
\,\overline{\xi_{\hat{n}^\mu} (k_\mu)}\,\Biggr] dk^4\Biggr\} d^3
{\bfm{k}}\,,}
\end{array}
\eqno{\rm (A.II.1A)}
$$

$$
\begin{array}{l}
{\ds \mathop{\Delta}\limits_{(a)} (n,t; \widehat{n},\widehat{t};
\mu ) := (2\pi)^{-1} \int\limits_{\R^3} \Biggr\{ \Biggr[
\,\prod\limits_{j=1}^{3} \xi_{n^j}(k_j) \,\overline{\xi_{\hat{n}^j}
(k_j)}\Biggr] } \\[0.5cm]
{\ds \Biggr[\,\int\limits_{C_{(a)}} (\eta^{\alpha\beta} k_\alpha
k_\beta + \mu^2)^{-1} \exp [ik_4(t-\widehat{t}\,)]\,dk^4\Biggr]
\Biggr\} d^3 {\bfm{k}}\,. }
\end{array}
\eqno{\rm (A.II.1B)}
$$
\vskip-2.2cm

$$ \begin{minipage}[t]{15cm}
\beginpicture
\setcoordinatesystem units <.75truecm,.75truecm>
\setplotarea x from 0 to 15, y from 7.5 to -9
\setlinear
\plot 0 0 14 0 /
\plot 7 4 7 -4 /
\put {\vector(1,0){4}} [Bl] at 14 0
\put {\vector(0,1){4}} [Bl] at 7 4

\setquadratic
\circulararc 360 degrees from 2 0 center at 2.5 0
\circulararc 360 degrees from 12 0 center at 11.5 0
\ellipticalarc axes ratio 4:1 360 degrees from 13.5 0 center at 7 0

\put {\circle*{3.25}} [Bl] at 2.5 0
\put {\circle*{3.25}} [Bl] at 11.5 0
\put {\vector(-3,4){4}} [Bl] at 2.95 0.215
\put {\vector(-3,4){4}} [Bl] at 11.95 0.215
\put {\vector(1,0){4}} [Bl] at 9.7 -1.475

\put {\vector(1,0){4}} [Bl] at 3  -2
\put {\vector(1,0){4}} [Bl] at 11 2

\setquadratic
\plot 5 -2  5.3 -1.95 5.5 -1.8 /
\plot 9 2  8.7 1.95  8.5 1.8 /

\setlinear
\plot 0 -2  5 -2 /
\plot 9 2   14 2 /
\plot 5.5 -1.8  8.5 1.8 /

\put {$-\omega$} at 2.5 -0.2
\put {$\omega $} at 11.5 -0.2
\put {$C_{-}$} at 3.3 -0.5
\put {$C_{+}$} at 12.3 -0.5
\put {$C$} at 10 -1.8
\put {$C_{F}$} at 3 -2.4
\put {$C_{F}$} at 11 2.4
\endpicture
\end{minipage} $$
\vskip1.8cm

\centerline{{\bf FIG.\,1} \quad The complex $k^4$-plane.}
\vskip1cm

\noindent
(Note that in our signature $k^4 = -k_4$.) The Green's functions
involving the closed contours in Fig.\,1 are called homogeneous,
whereas the Green's functions involving the open contours are called
inhomogeneous. Assuming appropriate uniform convergences of the
improper integrals in (A.II.1A,B) and using the equation (A.I.18),
we derive \vspace*{0.1cm}that
$$
\begin{array}{l}
\eta^{\mu\nu} \Delta_\mu^{\scsc\#} \Delta_\nu^{\scsc\#} \Delta_{(a)}
(n; \widehat{n}; \mu) - \mu^2 \Delta_{(a)} (n; \widehat{n}; \mu) \\[0.2cm]
{\ds = \quad \;\; - \int_{\R^3} \left\{ \int_{C(a)} \left[
\,\prod\limits_{\mu=1}^{4} \xi_{n^\mu} (k_\mu)
\overline{\xi}_{\hat{n}^\mu} (k_\mu)\right] dk^4\right\} d^3{\bfm{k}}
} \\[0.6cm]
= \; \left\{ \begin{array}{r}
{\ds - \sum\limits_{\mu=1}^{4} \delta_{n^\mu \widehat{n}^\mu} =:
-\delta_{n\widehat{n}}^{4} \quad \mbox{for the inhomogeneous,} }\\[-0.2cm]
\;\,\quad 0 \hfill \mbox{for the homogeneous;}
\end{array} \right.
\end{array}
\eqno{\rm (A.II.2A)}
$$
\vskip0.4cm

$$
\hspace*{-0.25cm}\begin{array}{l}
\delta^{jl} \Delta_j^{\scsc\#} \Delta_l^{\scsc\#} \Delta_{(a)}
({\bfm{n}},t; \widehat{\bfm{n}},\widehat{t}; \mu) - \partial_t^2
\Delta_{(a)} ({\bfm{n}},t; \widehat{\bfm{n}},\widehat{t}; \mu) \\[0.2cm]
\phantom{=\hspace*{0.005cm}} - \mu^2 \Delta_{(a)}
({\bfm{n}},t; \widehat{\bfm{n}},\widehat{t}; \mu ) \\[0.3cm]
{\ds = -(2\pi)^{-1} - \int_{\R^3} \Biggr\{ \Biggr[
\,\prod\limits_{j=1}^{3} \xi_{n^j} (k_j) \,\overline{\xi}_{\hat{n}^j}
(k_j)\Biggr] } \\[0.3cm]
{\ds \phantom{ = } \,\Biggr[ \int_{C(a)} \exp (ik_4(t-\widehat{t}\,))
dk^4 \Biggr] \Biggr\} d^3 {\bfm{k}} } \\[0.5cm]
{\ds = -\!\sum\limits_{j=1}^{3} \delta_{n^j \hat{n}^j} \delta
(t\!-\!\widehat{t}\,) } =: \left\{\!\!\!\begin{array}{r}
-\delta_{{\bfm{n}}\hat{\bfm{n}}}^{3} \delta (t\!-\!\widehat{t}\,)\;\;
\mbox{for the inhomogeneous,} \\[0.2cm]
\phantom{- } 0 \hfill \mbox{for the homogeneous;}\;\;\;
\end{array} \right.
\end{array}
\eqno{\rm (A.II.2B)}
$$
\vskip0.6cm

Consider the particular homogeneous Green's function $\Delta
({\bfm{n}},t; \widehat{\bfm{n}},\widehat{t}; \mu)$ in (A.II.1B) which is
associated with the closed contour $C$ in Fig.\,1. Performing the
closed contour integration in the complex $k^4$-plane, we
\vspace*{0.1cm}obtain
$$
\hspace*{-0.3cm}\begin{array}{l}
{\ds \Delta ({\bfm{n}},t; \widehat{\bfm{n}},\widehat{t}; \mu) =
-\!\!\int_{\R^3}\!\!\Biggr[\prod\limits_{j=1}^{3} \xi_{n^j} (k_j)
\,\overline{\xi}_{\hat{n}^j} (k_j)\Biggr]\!\big[\omega^{-1} \sin
(\omega(t\!-\!\widehat{t}\,))\big]\,d^3 {\bfm{k}}, } \\[0.6cm]
\Delta ({\bfm{n}},t; \widehat{\bfm{n}},\widehat{t}; \mu)
\equiv 0 \quad\mbox{for}\quad {\bfm{n}}\neq \widehat{{\bfm{n}}}\,,
\\[0.4cm]
\big[ \partial_t \Delta ({\bfm{n}},t; \widehat{\bfm{n}},\widehat{t};
\mu)\big]_{|\hat{t}=t} = -\delta_{{\bfm{n}}\hat{\bfm{n}}}^3\,,
\\[0.5cm]
\big[ \partial_t \partial_t \Delta ({\bfm{n}},t; \widehat{\bfm{n}},
\widehat{t}; \mu)\big]_{|\hat{t}=t} \equiv  0\,.
\end{array}
\eqno{\rm (A.II.3B)}
$$
\vskip0.3cm

\noindent
There exist the following relations among the Green's functions:

$$
\hspace*{-0.4cm}\begin{array}{l}
\Delta ({\bfm{n}},t; \widehat{\bfm{n}},\widehat{t}; \mu) =
\Delta_{+} ({\bfm{n}},t; \widehat{\bfm{n}},\widehat{t}; \mu)
+ \Delta_{-} ({\bfm{n}},t; \widehat{\bfm{n}},\widehat{t}; \mu)\,,
\\[0.2cm]
\Delta_{-} ({\bfm{n}},t; \widehat{\bfm{n}},\widehat{t}; \mu)
= \overline{\Delta_{+} ({\bfm{n}},t; \widehat{\bfm{n}},\widehat{t};
\mu)}\,, \\[0.2cm]
\Delta (\widehat{\bfm{n}},\widehat{t}; {\bfm{n}}, t; \mu)
= -\Delta ({\bfm{n}},t; \widehat{\bfm{n}}, \widehat{t}; \mu)\,,
\\[0.2cm]
\Delta_F ({\bfm{n}},t; \widehat{\bfm{n}}, \widehat{t}; \mu) := -\theta
(t-\widehat{t}\,) \Delta_{+} ({\bfm{n}},t; \widehat{\bfm{n}}, \widehat{t};
\mu) \\[0.1cm]
\;\qquad\qquad\qquad\qquad + \theta (\widehat{t}-t) \Delta_{-}
({\bfm{n}},t; \widehat{\bfm{n}}, \widehat{t}; \mu)\,, \\[0.2cm]
\Delta_{F} (\widehat{\bfm{n}},\widehat{t}; {\bfm{n}}, t; \mu)
= \Delta_{F} ({\bfm{n}},t; \widehat{\bfm{n}}, \widehat{t}; \mu)\,,
\\[0.2cm]
{\ds \theta (t) := \left(\frac{1}{2}\right) (1+t / |t|)
\quad\mbox{for}\quad t\neq 0\,. }
\end{array}
\eqno{\rm (A.II.4B)}
$$
\vskip0.2cm

We define the Green's functions for the massless case $\mu = 0$ by:
$$
D_{(a)} (n;\widehat{n}) := \Delta_{(a)} (n;\widehat{n};0)\,,
\eqno{\rm (A.II.5A)}
$$
$$
D_{(a)} ({\bfm{n}},t; \widehat{\bfm{n}},\widehat{t}) := \Delta_{(a)}
({\bfm{n}},t; \widehat{\bfm{n}},\widehat{t};0)\,.
\eqno{\rm (A.II.5B)}
$$
\vskip0.2cm

In the second quantization of the spin-$1/2$ fields, we encounter
the $4\times~4$ matrix-valued Green's functions [2] $S_{(a)} (n;
\widehat{n}; m)$ and $S_{(a)} ({\bfm{n}},t; \widehat{\bfm{n}},
\widehat{t}; m).$ These are defined by:
$$
\begin{array}{l}
S_{(a)} (n; \widehat{n}; m) := (\gamma^\mu \Delta_\mu^{\scsc\#}
-m\,{\mathrm{I}}) \Delta_{(a)} (n;\widehat{n}) \\[0.2cm]
{\ds = \int\limits_{\R^3} \Biggr\{ \int\limits_{C_{(a)}}
(\eta^{\alpha\beta} p_\alpha p_\beta + m^2)^{-1} (i\gamma^\mu p_\mu
- m\,{\mathrm{I}})} \\[0.7cm]
\phantom{ = } {\ds \Biggr[\,\prod\limits_{\nu=1}^{4} \xi_{n^\nu}
(p_\nu) \,\overline{\xi_{\hat{n}^\nu}} (p_\nu)\Biggr] dp^4\Biggr\}
d^3 {\bfm{p}}\,, }
\end{array}
\eqno{\rm (A.II.6A)}
$$
\vskip0.3cm

$$
\begin{array}{l}
S_{(a)} ({\bfm{n}},t; \widehat{\bfm{n}},\widehat{t}; m) := (\gamma^j
\Delta_j^{\scsc\#} + \gamma^4 \partial_t -m \,{\mathrm{I}}) \Delta_{(a)}
({\bfm{n}},t; \widehat{\bfm{n}},\widehat{t}; m) \\[0.2cm]
{\ds = (2\pi)^{-1} \int\limits_{\R^3} \Biggr\{ \int\limits_{C_{(a)}}
(\eta^{\alpha\beta} p_\alpha p_\beta + m^2)^{-1} (i\gamma^j p_j
+ i\gamma^4p_4 - m \,{\mathrm{I}}) } \\[0.6cm]
\phantom{ = } {\ds \Biggr[\,\prod\limits_{b=1}^{3} \xi_{n^b}
(p_b) \,\overline{\xi_{\hat{n}^b}} (p_b)\Biggr]
\exp \,(ip_4(t-\widehat{t}\,))\,dp^4 \Bigg\} d^3 {\bfm{p}}\,. }
\end{array}
\eqno{\rm (A.II.6B)}
$$
\medskip

\noindent
Here, $m>0$ is the mass parameter, $\gamma^\mu$ are Dirac matrices,
and $C_{(a)}$ are contours in the complex $p^4$-plane (exactly similar
to those in Fig.\,1).

We can prove that
$$
(\gamma^\mu \Delta_\mu^{\scsc\#} + m \,{\mathrm{I}}) S_{(a)} (n;
\widehat{n}; m) = \left\{\!\!\begin{array}{cl}
-\delta_{n\hat{n}}^{4}\,{\mathrm{I}} & \!\!\mbox{for the inhomogeneous,}
\\[0.15cm]
0 & \!\!$\mbox{for the homogeneous.}$ \end{array} \right.
\eqno{\rm (A.II.7A)}
$$

$$
\begin{array}{l}
(\gamma^j \Delta_j^{\scsc\#} + \gamma^4 \partial_t + m \,{\mathrm{I}})
S_{(a)} ({\bfm{n}},t; \widehat{\bfm{n}},\widehat{t}; m) \\[0.2cm]
= \left\{
\begin{array}{cl}
-\delta_{{\bfm{n}}\hat{\bfm{n}}}^{3} \delta (t-\widehat{t}\,)
\,{\mathrm{I}} & \mbox{for the inhomogeneous,} \\[0.15cm]
0 & $\mbox{for the homogeneous.}$ \end{array} \right.
\end{array}
\eqno{\rm (A.II.7B)}
$$
\vskip0.3cm

There exist linear relationships
$$
S(n;\widehat{n};m) = S_{+}(n;\widehat{n};m) + S_{-}(n;\widehat{n};m)\,,
\eqno{\rm (A.II.8A)}
$$
$$
S({\bfm{n}},t;\widehat{{\bfm{n}}},\widehat{t};m) =
S_{+}({\bfm{n}},t;\widehat{{\bfm{n}}},\widehat{t} ;m)
+ S_{-}({\bfm{n}},t; \widehat{{\bfm{n}}},\widehat{t};
m)\vspace*{0.2cm}\,,
\eqno{\rm (A.II.8B)}
$$
$$
\begin{array}{rcl}
S_F({\bfm{n}},t;\widehat{\bfm{n}},\widehat{t};m) &\!\!\!:=\!\!\!&
-\theta (t-\widehat{t}) S_{+} ({\bfm{n}},t;\widehat{\bfm{n}},
\widehat{t} ;m) \\[0.2cm]
&& + \theta (t-\widehat{t}) S_{-}({\bfm{n}},t;
\widehat{\bfm{n}},\widehat{t}; m)\,.
\end{array}
\eqno{\rm (A.II.8C)}
$$
\vskip1.2cm

\section*{Appendix III: \ Total momentum, energy, and charge of the
scalar field}

We shall compute here the total momentum $P_j$, the total energy
$H=-P_4$, and the total charge $Q$ from the non-hermitian scalar field
operator $\phi ({\bfm{n}},t)$ in the equations (7Bi,ii,iii). We have
from (8B),
$$
\begin{array}{rcl}
\phi ({\bfm{n}},t) &=& {\ds \int\limits_{\R^3} [2\omega
({\bfm{k}})]^{-1/2} \Biggr\{ a({\bfm{k}}) \Biggr[\,\prod\limits_{j=1}^3
\xi_{n^j} (k_j)\Biggr] e^{-i\omega t} } \\[0.5cm]
&& + {\ds b^{\dagger} ({\bfm{k}}) \Biggr[\,\prod\limits_{j=1}^3
\,\overline{\xi_{n^j}(k_j)}\Biggr] e^{i\omega t}\Biggr\} d^3
{\bfm{k}}\,, } \\[0.5cm]
\phi^{\dagger} ({\bfm{n}},t) &=& {\ds \int\limits_{\R^3} [2\omega
({\bfm{k}})]^{-1/2} \Biggr\{ a^{\dagger}({\bfm{k}}) \Biggr[
\,\prod\limits_{j=1}^3 \overline{\xi_{n^j} (k_j)}\Biggr] e^{i\omega t}
} \\[0.5cm]
&& + {\ds  b ({\bfm{k}}) \Biggr[\,\prod\limits_{j=1}^3
\xi_{n^j}(k_j)\Biggr] e^{-i\omega t}\Biggr\} d^3 {\bfm{k}}\,,}
\\[0.5cm]
\Delta_j^{\scsc\#} \phi ({\bfm{n}},t)_{|t=0} &=& {\ds i
\int\limits_{\R^3} k_j [2\omega ({\bfm{k}})]^{-1/2} \Biggr\{
a({\bfm{k}}) \Biggr[\,\prod\limits_{j=1}^3 \xi_{n^j} (k_j)\Biggr]
} \\[0.5cm]
&& - {\ds b^{\dagger} ({\bfm{k}}) \Biggr[\,\prod\limits_{j=1}^3
\,\overline{\xi_{n^j}} (k_j)\Biggr] \Biggr\} d^3 {\bfm{k}}\,,}
\\[0.5cm]
\partial_t \phi ({\bfm{n}},t)_{|t=0} &=& {\ds -i \int\limits_{\R^3}
\omega (\widehat{\bfm{k}}) [2\omega (\widehat{\bfm{k}})]^{-1/2}
\Biggr\{ a(\widehat{\bfm{k}}) \Biggr[ \prod\limits_{\ell=1}^3
\xi_{n^\ell} (\widehat{k}_\ell) } \\[0.5cm]
&& - {\ds b^{\dagger} (\widehat{\bfm{k}})
\prod\limits_{\ell=1}^3 \,\overline{\xi_{n^\ell}(\widehat{k}_\ell)}
\Biggr] \Biggr\} d^3 \widehat{\bfm{k}}\,. }
\end{array}
\eqno{\rm (A.III.1)}
$$
\medskip

We shall first compute the total charge $Q$ from (7Biii) and
(A.III.1) for the sake of simplicity.
$$
\begin{array}{l}
{\ds Q = ie \sum\limits_{{\bfm{n}}=0}^{\infty{(3)}} \big[\phi^{\dagger}
({\bfm{n}},t) \partial_t \phi ({\bfm{n}},t)\big]_{|t=0} +
{\mathrm{(h.c.)}} } \\[0.7cm]
{\ds = (e/2)\!\sum\limits_{{\bfm{n}}=0}^{\infty{(3)}}\,\int\limits_{\R^3}
\int\limits_{\R^3} (\widehat{\omega} / \omega )^{1/2} \left\{\left[
a^{\dagger} \widehat{a} \left( \prod\limits_{j} \prod\limits_{\ell}
\overline{\xi_{n^j}} \,\widehat{\xi}_{n^\ell}\right)
\!\!-\!b \widehat{b}^{\dagger} \left( \prod\limits_{j}
\prod\limits_{\ell} \xi_{n^j} \overline{\widehat{\xi}_{n^\ell}}\right)
\right] \right. } \\[0.7cm]
{\ds \left. + \left[ b\widehat{a} \left(\prod\limits_{j}
\prod\limits_{\ell} \xi_{n^j} \widehat{\xi}_{n^\ell}\right) - a^{\dagger}
\widehat{b}^{\dagger} \left( \prod\limits_{j} \prod\limits_{\ell}
\overline{\xi_{n^j}} \,\overline{\widehat{\xi}_{n^\ell}}\right)\right]
\right\} + {\mathrm{(h.c.)}}\,. }
\end{array}
$$
\vskip0.2cm

\noindent
Carrying out the triple sum $\sum\limits_{{\bfm{n}}}^{\infty}{}^{(3)}$
with help of the completeness relations (A.I.19) and (A.I.20), we
\vspace*{0.2cm}obtain
$$
\begin{array}{rcl}
Q &=& {\ds (e/2) \int\limits_{\R^3} \int\limits_{\R^3} (\widehat{\omega}/
\omega)^{1/2} \big[a^{\dagger} \widehat{a} - b \widehat{b}^{\dagger})
\,\delta^3 ({\bfm{k}}-\widehat{\bfm{k}}) } \\[0.6cm]
&& + (b\widehat{a} - a^{\dagger} \widehat{b}^{\dagger}) \delta^3 ({\bfm{k}}
+ \widehat{\bfm{k}})\big]\,d^3 {\bfm{k}}\,d^3 \widehat{\bfm{k}}
+ {\mathrm{(h.c.)}} \\[0.4cm]
&=& {\ds (e/2) \int\limits_{\R^3} \Big\{\big[ a^{\dagger} ({\bfm{k}})
a ({\bfm{k}}) - b({\bfm{k}}) b^{\dagger} ({\bfm{k}})\big] } \\[0.6cm]
&& + \big[b ({\bfm{k}}) a(-{\bfm{k}})
- a^{\dagger} ({\bfm{k}})
b^{\dagger} (-{\bfm{k}})\big]\Big\}\,d^3 {\bfm{k}}
+ {\mathrm{(h.c.)}} \\[0.4cm]
&=& {\ds e \int\limits_{\R^3} \big[ a^{\dagger} ({\bfm{k}})
a ({\bfm{k}}) - b({\bfm{k}}) b^{\dagger} ({\bfm{k}})\big]
\,d^3 {\bfm{k}}\,. }
\end{array}
\eqno{\rm (A.III.2)}
$$
\vskip0.2cm

\noindent
Using the definitions (11) and the commutators (10), we get
$$
Q = e \int\limits_{\R^3} [N^{+} ({\bfm{k}}) - N^{-} ({\bfm{k}})
- \delta^{3} ({\bfm{0}}) \,{\mathrm{I}} ({\bfm{k}})] \,d^3
{\bfm{k}}\,.
$$
The last divergent term in the above equation could have been avoided
by modifying [10] the Lagrangian (3). Thus we derive the equation
(13iii).

Now, we shall compute the total momentum components $P_j$ from
(A.III.1), (7Bi), and (10). The result is

$$
\begin{array}{l}
P_j = {\ds - \sum\limits_{{\bfm{n}}=0}^{\infty{(3)}} \big[
\partial_t \phi^{\dagger} ({\bfm{n}},t) \cdot \Delta_j^{\scsc\#}
\phi ({\bfm{n}},t)\big]_{|t=0} + {\mathrm{(h.c.)}} } \\[0.6cm]
= {\ds (1/2) \sum\limits_{{\bfm{n}}=0}^{\infty}\;\int\limits_{\R^3}
\int\limits_{\R^3} k_j (\widehat{\omega} / \omega )^{1/2} } \\[0.6cm]
{\ds \Biggr\{\Biggr[\widehat{a}^{\dagger} a \Biggr(\prod\limits_{\ell}
\,\prod\limits_{j} \overline{\widehat{\xi}_{n^\ell}} \,\xi_{n^j}\Biggr)
+ \widehat{b} b^{\dagger} \Biggr( \prod\limits_{\ell}\,
\prod\limits_{j} \widehat{\xi}_{n^\ell} \overline{\xi_{n^j}}\Biggr)
\Biggr] }
\end{array}
$$
$$
\begin{array}{l}
{\ds - \Biggr[ \widehat{a}^{\dagger} b^{\dagger} \Biggr(
\prod\limits_{\ell} \,\prod\limits_{j} \overline{\widehat{\xi}_{n^\ell}}
\,\overline{\xi_{n^j}}\Biggr)\!+\!\widehat{b} a \Biggr( \prod\limits_{\ell}
\prod\limits_{j} \widehat{\xi}_{n^\ell} \xi_{n^j}\Biggr)
\Biggr]\Biggr\} d^3 {\bfm{k}} d^3 \widehat{\bfm{k}} +
{\mathrm{(h.c.)}} } \\[0.6cm]
= {\ds (1/2) \int\limits_{\R^3} k_j \Big\{\big[ a^{\dagger}
({\bfm{k}}) a ({\bfm{k}}) - b({\bfm{k}}) b^{\dagger} ({\bfm{k}})\big]
- \big[a^{\dagger} (-{\bfm{k}}) b^{\dagger} ({\bfm{k}}) } \\[0.65cm]
+ b (-{\bfm{k}}) a ({\bfm{k}})\big]\Big\} d^3 {\bfm{k}}
+ {\mathrm{(h.c.)}} \\[0.25cm]
= {\ds \int\limits_{\R^3} k_j \big[ a^{\dagger}
({\bfm{k}}) a ({\bfm{k}}) + b^{\dagger} ({\bfm{k}}) b ({\bfm{k}})
+ \delta^3 ({\bfm{0}}) \,{\mathrm{I}} ({\bfm{k}}) \big]
d^3 {\bfm{k}} }
\end{array}
\eqno\raisebox{-9.5ex}{\rm (A.III.3)}
$$
\vskip0.3cm

\noindent
Here, we put $\delta^3 ({\bfm{0}}) \int\limits_{\R^3} k_j
\,{\mathrm{I}} ({\bfm{k}}) d^3 {\bfm{k}} = 0$ in the sense of the
Cauchy-Principal-Value and thus derive the equation (13i).

{F}inally, we calculate the total energy $H$ by the equations (A.III.1),
(7Bii), (10), and (11). We obtain
$$
\begin{array}{l}
\hspace*{-0.45cm}H = -P_4 \\[0.3cm]
= {\ds \sum\limits_{{\bfm{n}}=0}^{\infty{(3)}} \Big[
\delta^{ab} (\Delta_a^{\scsc\#} \phi^{\dagger} \cdot \Delta_b^{\scsc\#}
\phi ) + \partial_t \phi^{\dagger} \cdot \partial_t \phi }
+ \mu^2 \phi^{\dagger} ({\bfm{n}},t) \phi ({\bfm{n},t)}\Big]_{|t=0}
\\[0.5cm]
= {\ds (1/2) \sum\limits_{{\bfm{n}}=0}^{\infty}
\;\int\limits_{\R^3}
\int\limits_{\R^3} (\omega \widehat{\omega})^{-1/2} \Big[
(\delta^{ab} k_a \widehat{k}_b + \omega \widehat{\omega}
+ \mu^2) } \\[0.7cm]
\phantom{ =\; } (a^{\dagger} \widehat{a} + b \widehat{b}^{\dagger})\,
\delta^3 ({\bfm{k}}-\widehat{\bfm{k}})  \\[0.3cm]
\phantom{ = } -(\delta^{ab} k_a \widehat{k}_b + \omega \widehat{\omega}
- \mu^2) (a^{\dagger} \widehat{b}^{\dagger} + b \widehat{a})\, \delta^3
({\bfm{k}}+\widehat{\bfm{k}})\Big]\,d^3 {\bfm{k}}\,d^3 \widehat{\bfm{k}}
\\[0.4cm]
= {\ds \int\limits_{\R^3} \omega ({\bfm{k}}) \Big[a^{\dagger}
({\bfm{k}}) a ({\bfm{k}}) + b({\bfm{k}}) b^{\dagger} ({\bfm{k}})\Big]
\,d^3 {\bfm{k}}\,. }
\end{array}
\eqno{\rm (A.III.4)}
$$
\vskip0.2cm

By the above equation, the commutation relation (10), and the number
operators in (11), the total energy in (13ii) is obtained.

\newpage

\thispagestyle{empty}

\ \\[-2cm]

$$ \begin{minipage}[t]{15cm}
\beginpicture
\setcoordinatesystem units <0.9truecm,0.9truecm>
\setplotarea x from 0 to 15, y from 9 to -9
\setlinear
\plot 0 0 14 0 /
\plot 7 4 7 -4 /
\put {\vector(1,0){4}} [Bl] at 14 0
\put {\vector(0,1){4}} [Bl] at 7 4

\setquadratic
\circulararc 360 degrees from 2 0 center at 2.5 0
\circulararc 360 degrees from 12 0 center at 11.5 0
\ellipticalarc axes ratio 4:1 360 degrees from 13.5 0 center at 7 0

\put {\circle*{3.25}} [Bl] at 2.5 0
\put {\circle*{3.25}} [Bl] at 11.5 0
\put {\vector(-3,4){4}} [Bl] at 2.95 0.215
\put {\vector(-3,4){4}} [Bl] at 11.95 0.215
\put {\vector(1,0){4}} [Bl] at 9.7 -1.475

\put {\vector(1,0){4}} [Bl] at 3  -2
\put {\vector(1,0){4}} [Bl] at 11 2

\setquadratic
\plot 5 -2  5.3 -1.95 5.5 -1.8 /
\plot 9 2  8.7 1.95  8.5 1.8 /

\setlinear
\plot 0 -2  5 -2 /
\plot 9 2   14 2 /
\plot 5.5 -1.8  8.5 1.8 /

\put {$-\omega$} at 2.5 -0.2
\put {$\omega $} at 11.5 -0.2
\put {$C_{-}$} at 3.3 -0.5
\put {$C_{+}$} at 12.3 -0.5
\put {$C$} at 10 -1.8
\put {$C_{F}$} at 3 -2.4
\put {$C_{F}$} at 11 2.4

\put {{\bf FIG. 1} \ \ The complex $k^4$-plane.} at 7 -8

\endpicture
\end{minipage} $$


\vskip1.2cm


\begin{thebibliography}{99}
\medskip
\bibitem{[1]}
A. Das, submitted to Nucl. Phys. B.

\bibitem{[2]}
A. Das and P. Smoczynski, Found. Phys. Letters 7 (1994) 21, 127.

\bibitem{[3]}
A. Das, Nuovo Cimento 18 (1960) 482; Z. Phys. C, 41 (1988) 505.

\bibitem{[4]}
F. Riesz and B. Sz.-Nagy, Functional Analysis (translated by L.F.
Boron), (F. Ungar Publ. Co., New York, 1965), 212; E.R. Lorch, Spectral
Theory (Oxford University Press, New York, 1962), 60.

\bibitem{[5]}
A. Das, The Special Theory of Relativity: A Mathematical Exposition
(Springer-Verlag, New York, Heidelberg etc., 1993), 134, 145, 157.

\bibitem{[6]}
Y. Choquet-Bruhat, C. Dewitt-Morette, and M. Dillard-Bleick,
Analysis, Manifolds and Physics (North-Holland Publ. Co., New York,
Amsterdam etc., 1977), 129.

\bibitem{[7]}
H. Muirhead, The Physics of Elementary Particles (Pergamon Press,
New York etc., 1966), 139, 151, 158.

\bibitem{[8]}
L.P. Eisenhart, Riemannian Geometry (Princeton Univ. Press, Princeton,
1966), 96.

\bibitem{[9]}
G. Sansone, Orthogonal Functions (translated by A.H. Diamond),
(Interscience Publ. Inc., New York, 1959), 303. \\
H. Bateman, Higher Transcendental Functions, Vol.\,II (McGraw-Hill
Book Co. Inc., New York etc., 1953), 193. (Our definition of the
Hermite polynomial coincides with Bateman's and differs from Sansone's
by a factor $(-1)^n.$)

\bibitem[{10}]
{}W. Heinsenberg, Z. f{\"u}r Phys. 90 (1934) 209.
\end{thebibliography}
\end{document}